\documentclass[aps,prl,twocolumn,groupedaddress]{revtex4-1}

\usepackage{graphicx}
\usepackage{amsfonts}
\usepackage{amssymb}         
\usepackage[squaren,thinspace]{SIunits}
\usepackage{onlyamsmath}
\usepackage{xcolor}

\begin{document}

\newcommand{\By}{$\times$}
\newcommand{\SqrtBy}[2]{$\sqrt{#1}$\kern0.2ex$\times$\kern-0.2ex$\sqrt{#2}$}
\newcommand{\Degree}{$^\circ$}
\newcommand{\DegreeC}{$^\circ$C}
\newcommand{\Ohmcm}{$\Omega\cdot$cm}

\title{Stabilizing isolated skyrmions at low magnetic fields exploiting vanishing magnetic anisotropy} 

\author{Marie Herv\'e$^{1}$, Bertrand Dup\'e$^{2,3}$, Rafael Lopes$^{4}$, Marie B\"ottcher$^{2}$, Maximiliano D. Martins$^4$,   Timofey Balashov$^{1}$, Lukas Gerhard$^{5}$, Jairo Sinova$^{2,6}$, Wulf Wulfhekel$^{1,5}$ \\
$^1$Physikalisches Institut, Karlsruhe Institute of Technology, 76131 Karlsruhe, Germany \\
$^2$Institut f\"ur Physik, Johannes Gutenberg Universit\"at Mainz, D-55099 Mainz, Germany \\
$^3$Institute of Theoretical Physics and Astrophysics, University of Kiel, 24098 Kiel, Germany \\
$^4$Centro de Desenvolvimento da Tecnologia Nuclear, 31270-901, Belo Horizonte, Brazil \\
$^5$Institute of Nanotechnology, Karlsruhe Institute of Technology,  76128 Karlsruhe, Germany \\
$^6$Institute of Physics, Academy of Sciences of the Czech Republic, Cukrovarnick\'{a} 10, 162 53 Praha 6 Czech Republic \\
Corresponding author: marie.herve@kit.edu, bertdupe@uni-mainz.de\\}

 \maketitle

{\bf 
Skyrmions are topologically protected non-collinear magnetic structures. Their stability and dynamics, arising from their topological character, have made them ideal information carriers e.g. in racetrack memories. The success of such a memory critically depends on the ability to stabilize and manipulate skyrmions at low magnetic fields. 
The driving force for skyrmion formation is the non-collinear Dzyaloshinskii-Moriya exchange interaction (DMI) originating from spin-orbit coupling (SOC). It competes with both the nearest neighbour Heisenberg exchange interaction and the magnetic anisotropy, which favour collinear states. While skyrmion lattices might evolve at vanishing magnetic fields, the formation of isolated skyrmions in ultra-thin films so far required the application of an external field which can be as high as several T. 
Here, we show that isolated skyrmions in a monolayer (ML) of Co epitaxially grown on a Ru(0001) substrate can be stabilized at magnetic fields as low as 100 mT. Even though SOC is weak in the 4d element Ru, a homochiral spin spiral ground state and isolated skyrmions could be detected and laterally resolved using a combination of tunneling and anisotropic tunneling magnetoresistance effect in spin-sensitive scanning tunneling microscopy (STM).
Density functional theory (DFT) calculations confirm these chiral magnetic textures, even though the stabilizing DMI interaction is weak. We find that the key factor is the absence of magnetocristalline anisotropy in this system which enables non-collinear states to evolve in spite of weak SOC, opening up a wide choice of materials beyond 5d elements.}

\newpage

Topological spin structures are attracting rising attention due to their inherent magnetic stability~\cite{Bocdanov-1994aa,nature05056,Bogdanov1989}. Among these structures, isolated skyrmions are of particular interest since they can be moved by currents of very low density~\cite{Jonietz1648,nnano.2013.29,Sampaio2013}. The stability of skyrmions is ultimately linked to the DMI, which is a relativistic effect ~\cite{Dzyaloshinsky1958,Dzyaloshinskii1964,PhysRev.120.91} that favours non-collinear spin structures of a unique sense of rotation of neighboring magnetic moments.
To display a DMI, the spatial inversion symmetry of the magnetic material needs to be broken either in the bulk by the crystal structure itself as in MnSi ~\cite{muhlbauer2009skyrmion}, in layered systems by asymmetric interfaces as in $[$Pt/Co/Ir$]_{\mathrm{n}}$~\cite{Moreau-Luchaire2016}, $[$Rh/Pd/2Fe/2Ir$]_{\mathrm{n}}$~\cite{Dupe2016b}, or in ultra-thin films as in Fe monolayers on a Ir(111) substrate~\cite{nphys2045}. 
Isolated skyrmions have been observed in a wide range of polycrystalline metallic films such as $[$Pt/Co/Ir$]_{\mathrm{n}}$~\cite{Moreau-Luchaire2016}, CoFeB/Ta ~\cite{Jiang2015a} or in systems consisting of dipolar coupled magnetic films, with each of the films having non-symmetric interfaces~\cite{2016arXiv161100647H, Pollard2017}. However, in these polycrystalline systems, skyrmion mobility is limited by pinning to the large amount of structural defects.
For epitaxial ultra-thin films, only few systems are known to stabilize magnetic skyrmions. Fe(1ML)/Ir(111) shows a skyrmion lattice as the ground state, but no isolated skyrmions. Only in Fe(3ML)/Ir(111) ~\cite{hsu2017} and Pd/Fe(1ML)/Ir(111) ~\cite{Romming2013}, isolated magnetic skyrmions were reported in the presence of large magnetic fields ($\approx$ 1 to \unit{3}{\tesla}).

\begin{figure*}
	\begin{center}[h]
		\includegraphics[width=17cm]{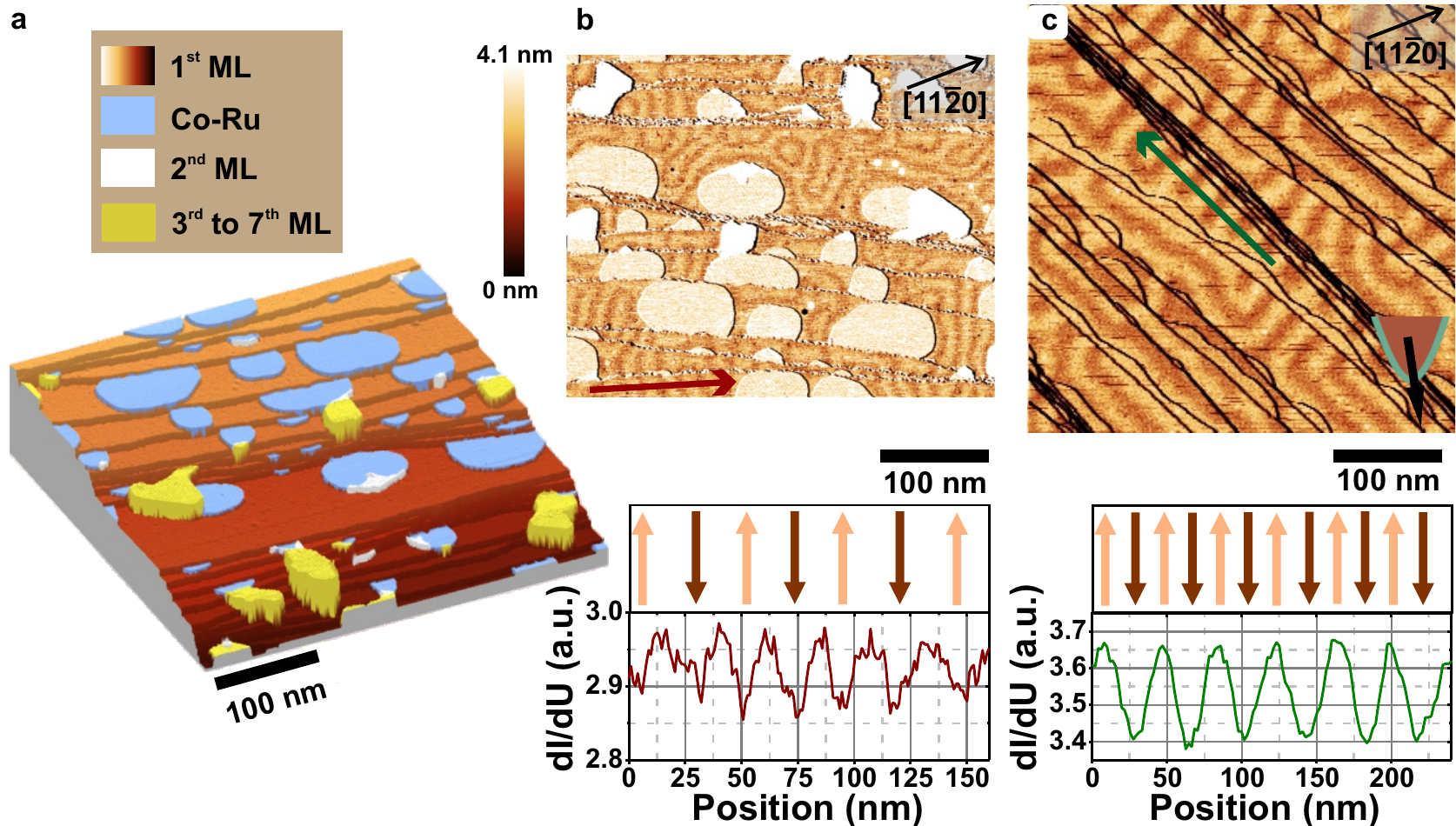}
	\end{center}
	\caption{\label{fig_1} {\bf Topography and magnetic structure of Co films on Ru(0001).}
		(a) STM topography of 1.1 ML Co ($I$=\unit{10}{\nano\ampere}, $U$=\unit{-320}{\milli\volt}). (b) Top: $dI/dU$ map of the same area as (a) taken with a W tip ($\Delta{U}_{rms}$=\unit{50}{\milli\volt}) revealing a periodic stripe pattern through TAMR effect. Bottom: line profile of $dI/dU$ plotted along the red arrow in the image and corresponding sketch of the magnetization. (c) Top: spin polarized $dI/dU$ map taken with a Cr coated W tip revealing the periodic magnetic stripe pattern through the TMR effect ($I$=\unit{1}{\nano\ampere}, $U$=\unit{-300}{\milli\volt}, $\Delta{U}_{rms}$=\unit{30}{\milli\volt}). The tip spin polarization was out-of-plane. Bottom:  Line profile of $dI/dU$ plotted along the green arrow in the image and corresponding sketch of the magnetization.}
\end{figure*}
A well established process to create skyrmions in thin films that display a chiral spin spiral ground state is to apply magnetic fields that cause the spirals to evolve into skyrmions \cite{Nandy2016}. The spin spiral ground state results from the competition between DMI, Heisenberg exchange and magnetic anisotropy energy (MAE) ~\cite{Dupe2014,Simon2014,Leonov2015a}. As has been shown in several theoretical works~\cite{Bocdanov-1994aa,Rohart2013}, a critical DMI $D_c$ is necessary to form spin spirals :          

\begin{equation}
D_c	\propto \sqrt[]{A k},
\end{equation}

\noindent where $A$ is the spin stiffness and $k$ the MAE constant.
One way to stabilize the non-collinear spin structure is to enlarge the DMI above $D_c$. This can be achieved with interfacing the magnetic atoms and 5d elements showing a large SOC as it was done in Fe/Ir(111) based structures. An alternative approach taken in this work is to reduce $D_c$ itself. $D_c$ can be lowered to arbitrarily small values when reducing the MAE to zero, rendering strong DMI an unnecessary criterion and opening up materials choice beyond 5d elements.

Here, we report on the creation of isolated skyrmions in 1 ML of Co on Ru(0001). This is the first reported ultra-thin film Co system deviating from a simple collinear ground state. 
By using a combination of the tunneling magnetoresistance (TMR) and tunneling anisotropic magnetoresistance (TAMR) in an STM  experiment, we were able to demonstrate that both the spin-spiral and the isolated skyrmion are homochiral. In contrast to previous approaches \cite{BodeNature,Romming2015}, this approach does not require large magnetic fields to be applied in different directions to proof a unique rotational sense. DFT calculations show that the stabilization mechanism of the isolated skyrmions differs from the ones reported in Fe/Ir(111) \cite{nphys2045} and Pd/Fe/Ir(111) \cite{Romming2013}, where the DMI is the leading energy term due to large SOC. In Co/Ru(0001), the magnetic exchange interaction is stiffer by a factor of 2 and the DMI smaller by a factor of 6 as compared with Pd/Fe/Ir(111). Nevertheless, chiral magnetic states are favoured due to the very weak MAE. This is the first experimental realization of this scenario initially introduced theoretically by Rohart {\it et al.}~\cite{Rohart2013}.

{\bf Magnetic ground state at zero magnetic field}

Figure \ref{fig_1}a shows the topography of a 1.1 ML film of Co deposited on a clean Ru(0001) substrate at $\approx$  \unit{300}{\celsius}. One ML is defined as one Co atom per substrate atom. At this deposition temperature, the first Co ML is known to grow pseudomorphically with hcp stacking forming a closed wetting layer on the  substrate \cite{growth}. In agreement to this, the STM topography shows Ru atomic terraces fully covered by one Co ML. The lower edges are decorated with islands of an alloy of Co and Ru (for details, see Supplementary Fig. S1).  The study of the structural and magnetic properties of these islands is beyond the scope of this communication and will be presented elsewhere. Few thicker islands (2-7 ML) are present due to the coverage being larger than 1 ML. These islands show a network of stacking faults as previously reported \cite{growth}.  

We performed spin-resolved STM to reveal the magnetic structure of the first ML.
For this, the spatial variation of the differential tunneling conductance ($dI/dU$) was mapped using a a bare tungsten, i.e. non-spin polarized tip (see Fig. \ref{fig_1}b) and a spin-polarized tip (see Fig. \ref{fig_1}c). Both maps were recorded at zero applied magnetic field. 
In Fig. \ref{fig_1}b, the alloy islands, as well as the thicker islands appear brighter than the ML due to their higher density of states.

On the ML, a periodic stripe modulation is observed with a periodicity of about 20 nm (see line profile along the red arrow in Fig. \ref{fig_1}b) when imaged using a non-magnetic tip and of about 40 nm when imaged using a spin-polarized tip with out-of plane polarization (see Fig. \ref{fig_1}c). 
Since the stripe patterns in both cases react on the application of a moderate magnetic field of several 100 mT, the contrast must be of magnetic origin (see Supplementary Fig. S2).
Several mechanisms may lead to stripe patterns in thin magnetic films. First of all, stripes can be stabilized by the magnetic dipolar energy. In the case of a single ML, however, the stripe periodicity is expected to diverge \cite{PhysRevB.75.014406}. Second, a DMI \cite{Dzyaloshinsky1958,Dzyaloshinskii1964,PhysRev.120.91} or a oscillatory long-range exchange interaction \cite{sanders} may induce non-collinear ground states consisting of spin spirals. While the DMI favours a unique rotational sense of the spin spirals, i.e. a chiral spin structure \cite{BodeNature}, the long range exchange interaction is symmetric and spin spirals of both rotational senses are degenerate.
In the spin-polarized $dI/dU$ map (Fig. \ref{fig_1}c), the local magnetization rotates by 180$^{\circ}$ between consecutive bright and dark stripes with the local sample magnetization pointing ``upward'' in the dark area (here ``upward'' is defines as antiparallel to the tip spin polarization) and ``downward'' in the bright area. A detailed demonstration of this is discussed in the Supplementary Fig. S2. 
When the spin structure is investigated with a bare W tip (Fig. \ref{fig_1}b), the local sample magnetization can still be sensed by the TAMR \cite{bode,TAMR,VanBergmann}. In systems with SOC, the electronic band structure depends on the axis of the magnetization orientation giving rise to changes of the local density of states (LDOS) at specific bias voltages, i.e. areas magnetized in-plane or out-of-plane may exhibit different LDOS. Thus, the experiment indicates that the local magnetization rotates by 90$^\circ$ between consecutive bright and dark stripes explaining the halved periodicity compared to spin-polarized measurements. 
In the line profiles in Fig. \ref{fig_1}b and c, the magnetic contrast varies continuously as a function of position, i.e. the stripes are not magnetic domains separated by sharp domain walls but the magnetization gradually rotates as a function of position in the form of a spin spiral. 
\begin{figure}[h]
	\begin{center}
		\includegraphics[width=8.5cm]{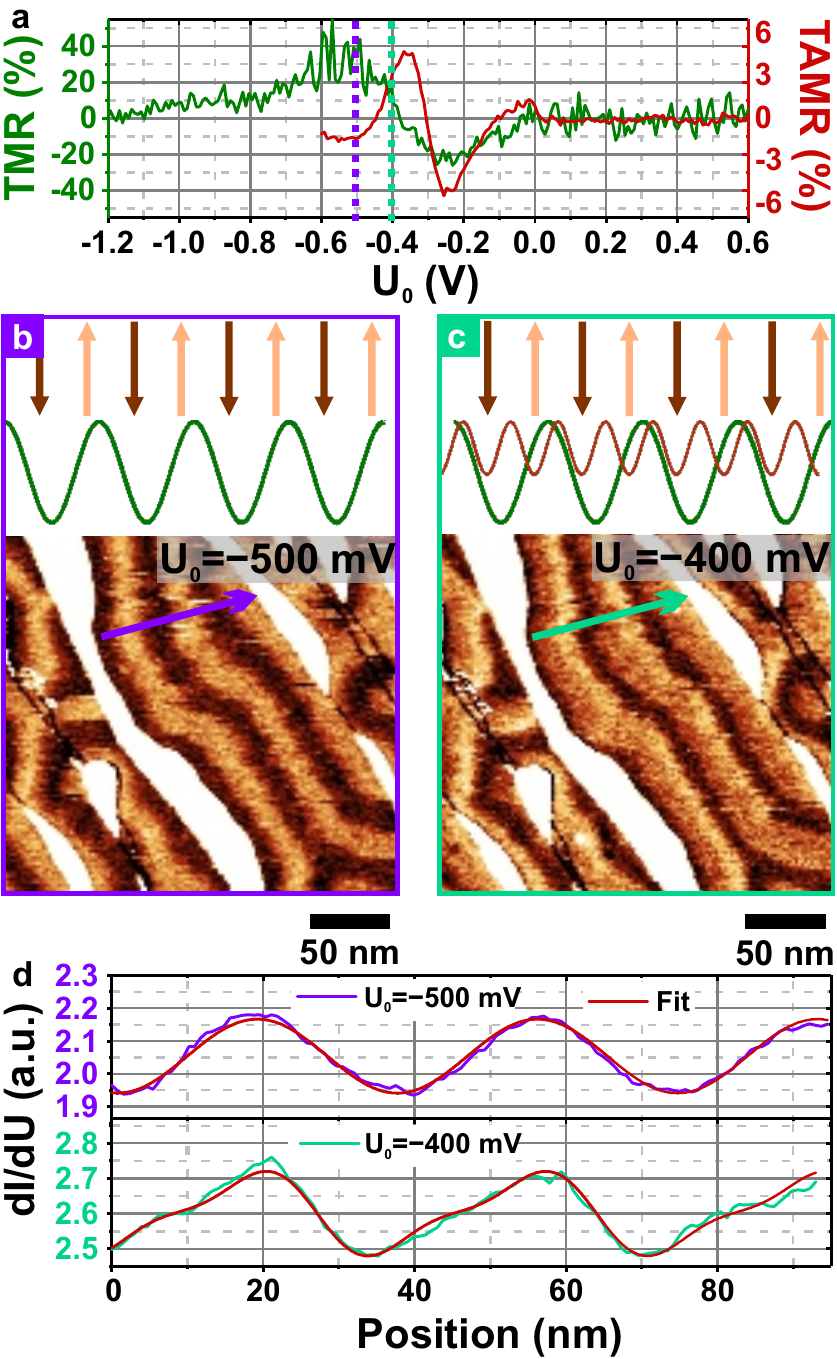}
	\end{center}
	\caption{\label{fig_3}
		{\bf Chirality of the spin spiral.}
		(a) Voltage dependency of the TMR (green curve - $\Delta{U}_{rms}$=\unit{20}{\milli\volt}) and TAMR (red curve - $\Delta{U}_{rms}$=\unit{10}{\milli\volt}). (b) and (c) bottom: $dI/dU$ spin polarized map taken at $U=$\unit{-500}{\milli\volt} (b) and \unit{-400}{\milli\volt} (c) ($I$=\unit{1}{\nano\ampere}, $\Delta{U}_{rms}$=\unit{40}{\milli\volt}, tip spin polarization was at 47$^{\circ}$ away from the normal direction). Top: sketches of the magnetization and the corresponding $dI/dU$ profiles perpendicular to the stripes due to TMR effect (green curve) and TAMR effect (red curve). (d) experimental $dI/dU$ profile plotted along the violet arrow in (b) and the green arrow in (c). Red curves superimposed on the experimental profiles are fits assuming a contribution of TMR only for the top panel and a combination between TMR and TAMR for the bottom panel.}
\end{figure}

While an oscillating long-range exchange interaction will result in randomly rotating transitions, the DMI favours a specific chiral rotational sense, i.e. a homochiral spin spiral. Using in-plane polarized tips, the first will show in-plane stripes with random sequence of contrast while the latter with alternating sequence of contrast \cite{bode2,Romming2015}. Further, the sign of the chirality can be achieved by a vectorial measurement of the spin polarization in large magnetic fields \cite{PhysRevB.95.060415}.
As the spin spirals observed in this work are strongly modified even with small fields (see Supplementary Fig. S2), we introduce an alternative method to investigate the chirality employing a combination of TAMR and TMR. 
The $dI/dU$ contrast due to TAMR recorded with a bare W tip as a function of bias voltage is shown in Figure \ref{fig_3}a (red curve). It sensitively depends on the bias voltage, is significant only in the range between $-420$ and \unit{-170}{\milli\volt} and displays a change of sign at \unit{-300}{\milli\volt}. As shown in the Supplementary (Fig. S3), the TAMR at \unit{-400}{\milli\volt} is positive, i.e. we observe a maximal signal for an in-plane local magnetization. Respectively, the $dI/dU$ signal is minimal when the local magnetization is out-of-plane. Note that the TAMR was determined from the contrast of 20 nm periodicity, i.e. between in- and out-of-plane magnetized areas.
We also measured the voltage dependence of the TMR (green curve - Fig. \ref{fig_3}a). This was determined with an out-of-plane spin-polarized tip studying the contrast with a 40 nm periodicity, i.e. comparing up- and down-magnetized areas. Thus, this excludes contributions of the TAMR (for more information see Supplementary Fig. S2).
The TMR contrast is significant for bias voltages between \unit{-800}{\milli\volt} and \unit{-50}{\milli\volt}. As shown in Fig. \ref{fig_3}a,  at \unit{-500}{\milli\volt} (see vertical violet line in Fig. \ref{fig_3}a), the TAMR is negligible ($\approx$ 1 \%) in comparison to the strong TMR ($\approx$ 30 \%). However, at \unit{-400}{\milli\volt} (see vertical green line in Fig. \ref{fig_3}a), the TAMR  ($\approx$ 3 \%) and TMR  ($\approx$ 10 \%) are of comparable amplitude. 
\begin{figure*}
	\begin{center}
		\includegraphics[width=13cm]{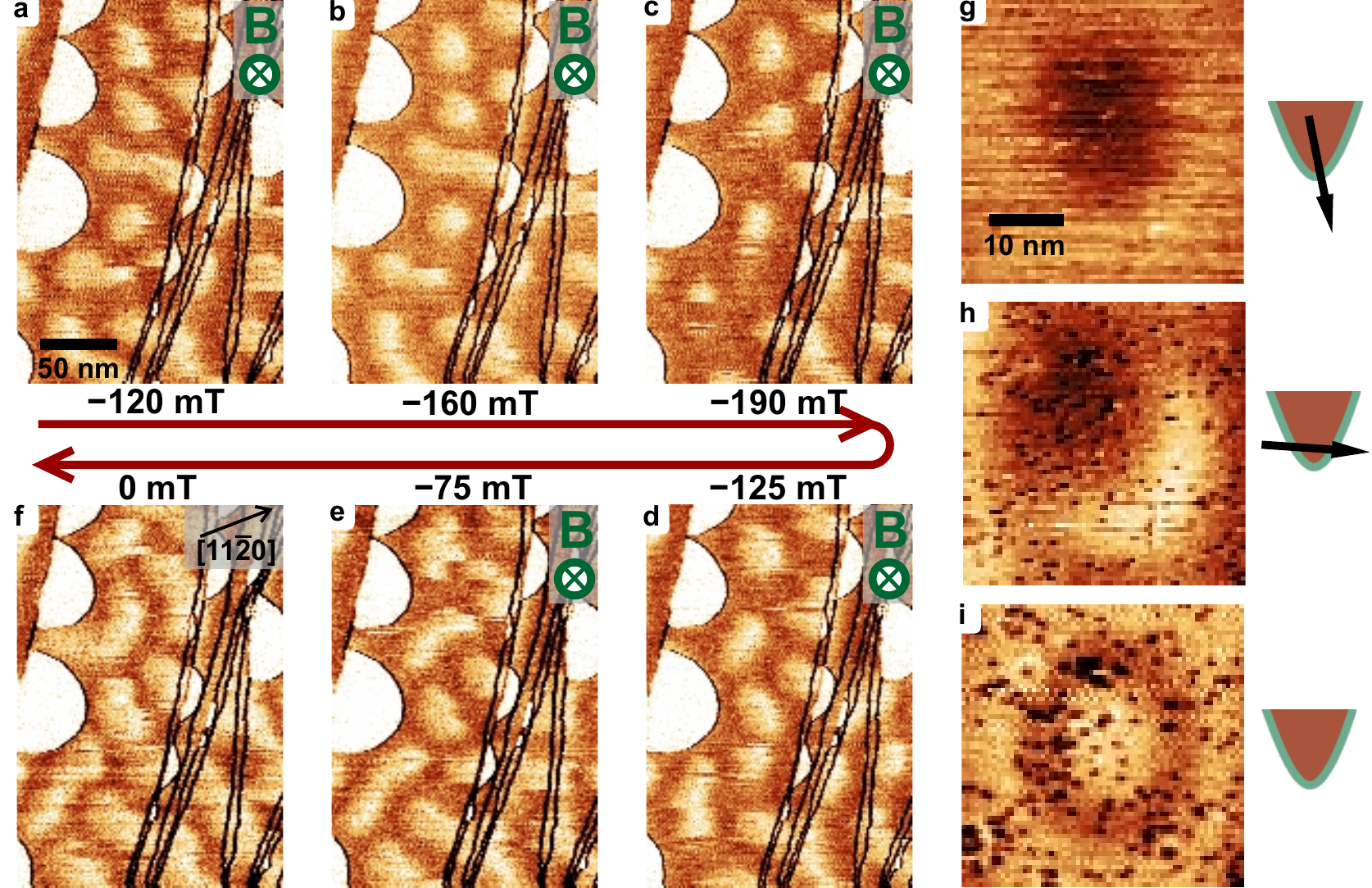}
	\end{center}
	\caption{\label{fig_2}
		{\bf Stabilization of magnetic skyrmions.}
		(a-f) Spin-polarized $dI/dU$ maps of the same area taken with out-of-plane magnetic fields as indicated ($I$=\unit{1}{\nano\ampere}, $U$=\unit{-400}{\milli\volt}, $\Delta{U}_{rms}$=\unit{40}{\milli\volt}). (g-i) Magnified $dI/dU$ maps recorded on the same skyrmion with an out-of-plane (g), in-plane (h) and non (i) spin-polarized tip ($I$=\unit{1}{\nano\ampere}, $U$=\unit{-400}{\milli\volt} for (g) and (h) and \unit{-200}{\milli\volt} for (i), $\Delta{U}_{rms}$=\unit{40}{\milli\volt} for (g) and (h) and \unit{30}{\milli\volt} for (i)).}
\end{figure*}

Fig. \ref{fig_3}b and c display two $dI/dU$ maps recorded on the same area of the sample with the same spin-polarized tip at two different bias voltages: $U$=-\unit{500}{\milli\volt} for Fig. \ref{fig_3}b and $U$=-\unit{400}{\milli\volt} for Fig. \ref{fig_3}c. As at \unit{-500}{\milli\volt}, the TMR dominates, the $dI/dU$ profile plotted along the axis perpendicular to the stripes shows a sinusoidal behaviour (violet curve in Fig. \ref{fig_3}d) that can be well fitted by a simple sine function (red curve) of \unit{37}{\nano\meter} periodicity. At \unit{-400}{\milli\volt}, the corresponding $dI/dU$ profile plotted in Fig. \ref{fig_3}d (green curve) shows a ``sawtooth''  periodic shape. At this bias voltage, both, TAMR and TMR are present and the signal is composed of a TMR contribution (green curve, sketch top panel Fig. \ref{fig_3}c), and a TAMR contribution (red curve, sketch top panel Fig. \ref{fig_3}c). The TAMR signal is independent of the direction of the tip spin polarization, while the TMR signal depends on the orientation of the magnetization of the tip. The measured combination of both signals thus contains information on the direction of the tip magnetization.  In order to quantitatively describe the signal, we fitted the data using the following expression :

\begin{equation}
\frac{d I}{ d U}(x)=A_1\sin (\frac{2 \pi}{\lambda}x+\phi_1)+A_2\sin (\frac{4 \pi}{\lambda}x+\phi_2).
\end{equation}

The first term corresponds to the TMR, the second to the TAMR signal. We fitted the experimental data using a periodicity of the spin structure of $\lambda$ of \unit{37}{\nano\meter}. Amplitudes ($A_{1}$, $A_{2}$) and phases ($\phi_{1}$, $\phi_{2}$) were the adjustable parameters. From the phase information, the tip spin polarization orientation was deduced to be at 47$^{\circ}$ away from the normal direction. By looking at the $dI/dU$ map shown in Fig. \ref{fig_3}c from left to right along the axis perpendicular to the stripes, one can see that the contrast everywhere in the image slowly rises from dark to bright, followed by a sharp transition back to dark. A non-chiral spin structure would show arbitrary changes of the rotational sense, which would lead to reversals of the sawtooth profile. As this is not the case, we conclude that a unique rotational sense is preferred in this system, i.e. the spin spiral is chiral rotating continuously by 360$^\circ$ per period and the DMI sets the sense of chirality.

{\bf Magnetic skyrmions}

Figures \ref{fig_2}a to f display six spin-polarized $dI/dU$ maps recorded consecutively with an out-of-plane spin-polarized tip at perpendicular magnetic fields as indicated. As already discussed above and in the Supplementary (Fig. S2), the magnetic structure is significantly modified even by modest fields. Dark areas expand with the applied out-of-plane field indicating a local orientation parallel to that of the field. Some of the remaining bright areas where the magnetization is antiparallel to the field tend to form circular dots of anti-aligned magnetization in an environment of magnetization aligned with the field. Upon further increase of the magnetic field (Fig. \ref{fig_2}a to c) some of these dots disappear when encountering structural defects such as step edges. When decreasing the magnetic field from -\unit{190}{\milli\tesla} to \unit{0}{\milli\tesla} (Fig. \ref{fig_2}c to f), the dots elongate to worm-like domains and finally to the spin-spiral structure. This behaviour has is analogous to the elliptical instability of bubble domains \cite{Bocdanov-1994aa,Leonov2015a}. After ramping the field down to zero, some circular dots remained stable. This behaviour is mostly found on narrow atomic terraces, as is discussed in the Supplementary (Fig. S4).

Fig. \ref{fig_2}g to i show three $dI/dU$ maps of such circular spin structures. As for Cr coated W tips, the spin polarization is extremely sensitive to the tip termination, slight voltage pulses were sufficient to switch the tip spin polarization from out-of-plane (Fig. \ref{fig_2}g) to in-plane (Fig. \ref{fig_2}h) to unpolarized (Fig. \ref{fig_2}i).

First, we focus on the results obtained with an unpolarized tip.  Fig. \ref{fig_2}i was recorded at -\unit{200}{\milli\volt}, where the TAMR is negative, i.e. in-plane magnetization give a lower $dI/dU$ signal. The map clearly shows a continuous transition of an out-of-plane magnetization in the center of the spin structure via an in-plane orientation (dark ring) back to an out-of-plane orientation with a similar rate as observed in the chiral spin spiral. Thus the structure does not represent a bubble domain with sharp domain walls but resembles a proper skyrmion. Fig. \ref{fig_2}g and h give information on the direction of rotation of the magnetization. The out-of-plane image indicates that the center of the skyrmion is magnetized in opposite direction of the surroundings and the in-plane image reveals a specific rotation sense of the spin structure. The contrast reveals two lobes: the bright and dark one correspond to a tip spin polarization parallel and antiparallel to the local sample magnetization respectively. Note that all observed circular spin structures display the same orientation of the in-plane contrast confirming the role of DMI (see Supplementary Fig. S4).
Further note that the two images were recorded at -400 mV, where also a small TAMR is present leading to a slight overshoot of the $dI/dU$ signal in Fig.  \ref{fig_2}g.
We can conclude that the topology of the observed spin structure is that of skyrmions, i.e. it exhibits a winding number of 1. This observation is in agreement with the unique chirality of the spin spiral discussed above.
\begin{figure}[h]
	\begin{center}
		\includegraphics[width=8.5cm]{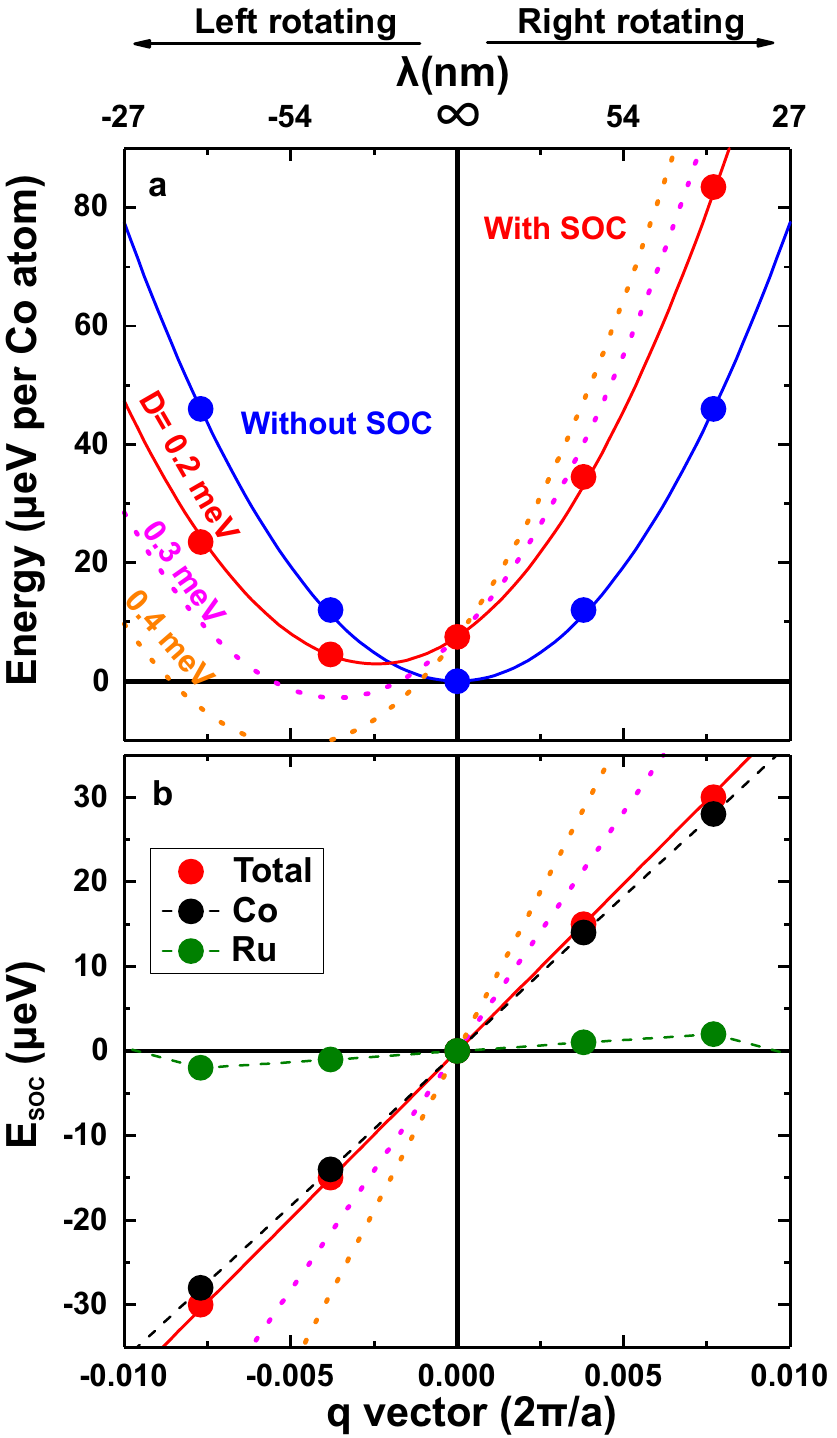}
	\end{center}
	\caption{\label{fig_T1}
		{\bf Spin spiral dispersion curve.}
		(a) Energy of a left and right rotating flat spin spiral as a function of the \textbf{q} vector (in cartesian coordinate) along the $\bar{\Gamma}-\bar{\mathrm{M}}$ direction. The continuous lines correspond to fits of the dispersions curve close to the $\bar{\Gamma}$-point ($\mathbf{q}=0$) without SOC (blue) and with SOC contribution (red). (b) Layer decomposition of the SOC contribution to the dispersion curve. The continuous line correspond to a fit of the SOC contribution on the effective DM interaction written in eq. \ref{DMI}. On both panels, the dashed magenta and orange lines correspond the dispersion curve with SOC contribution when the DMI is set to $D_1=0.3$~meV and $D_1=0.4$, respectively.}
\end{figure}

{\bf Theoretical model}

In order to understand the microscopic mechanism of skyrmion formation in Co/Ru(0001), we have performed DFT calculations using the FLEUR ab initio package~\cite{FLEUR}. FLEUR uses the full linearized augmented plane wave basis set (FLAPW) which has been successful in predicting the magnetic properties of 3d monolayers hybridized with 4d or 5d non-magnetic layers~\cite{nphys2045,Dupe2014,Dupe2016b,PhysRevB.79.094411}. In order to obtain the magnetic exchange interaction, we calculate the energy of spin-spirals~\cite{PhysRevB.69.024415} characterized by a constant angle between neighboring magnetic moments $\theta_i= \mathbf{q}\cdot\mathbf{r}_i$. Here $\mathbf{q}$ is the propagation direction of the spin spiral and $\mathbf{r}_i$ is the position of the $i^{\mathrm{th}}$ magnetic moment. In the case of Co/Ru(0001), the experimental angle between neighboring spins is rather small ($\theta=2.4^{\circ}$) which correspond to a propagation vector of $q=0.0067 \: 2\pi/a$. This justified to consider the spin spiral as a perturbation from the ferromagnetic (FM) state, i.e. to use the magnetic force theorem~\cite{PhysRevB.39.86}.

Figure~\ref{fig_T1}(a) shows the energy of the spin spiral with respect to the FM state with and without SOC for small $q$. As the angle between adjacent spins increases, the exchange energy rises. Without SOC, both right- and left-rotating states are degenerate (blue curve) and the exchange interaction favors a FM state ($q=0$). When SOC is taken into account, the degeneracy is lifted and an energy minimum is created for the left rotating spin spiral (red curve). Note that the SOC also changes the energy of the FM state due to the MAE. The $q=0$ state here reflects the energy of the FM state averaged over the different possible magnetization directions. It is 1.4 $\mu$eV/Co above the FM state with an in-plane magnetization.
 
To disentangle the different magnetic interactions, we have mapped our DFT result onto a Heisenberg Hamiltonian. The dispersion curves without SOC can be approximated by a first neighbor exchange interaction. In the case of Co/Ru(0001), we found $J_{\mathrm{eff}}=13.1$~meV/Co. This value is much larger than $J_{\mathrm{eff}}=-2.3$ and $J_{\mathrm{eff}}=4.4$~meV/Fe in the case of previous ultra-thin films where isolated skyrmions were found~\cite{Dupe2014,Simon2014}. 

The presence of isolated skyrmions can be explained by analysing the SOC contributions. On the one hand, the MAE is found to be in the range $\kappa_{\mathrm{FM}} \in [15,50]$~$\mu$eV/Co atom and favors an in-plane magnetization (offset of the dispersion $E(q)$ including SOC with respect to the $E=0$~axis). The MAE is one order of magnitude lower than for Pd/Fe/Ir(111) where $\kappa= \: \sim 1$~meV/Fe.
The presence of this small MAE was already suggested by previous works and explained by the structural strain in the Co monolayer ~\cite{SPLEEM}. On the other hand, the SOC favors a particular rotational sense, e.g. a left rotating spin spiral. To analyze this contribution, we have mapped the SOC contribution shown in Fig. \ref{fig_T1}b onto the effective DMI Hamiltonian:

\begin{equation}
	E_{\mathrm{DMI}}=\sum_{i,j} \mathbf{D}_{ij} \cdot \left( \mathbf{M}_i \times \mathbf{M}_j \right),
	\label{DMI}
\end{equation}

\noindent where $\mathbf{D}_{ij}$ contains both the direction and the amplitude of the DMI, $\mathbf{M}_i$ and $\mathbf{M}_j$ are unit vectors collinear to the magnetic moments at position $\mathbf{r}_i$ and $\mathbf{r}_j$, respectively. We calculate $D_{ij}=0.2$~meV/Co. 
It is much lower than the previously published values for Mn/W(001) ($D_{ij}=4.6$~meV/Mn)~\cite{bode2}, Pd/Fe/Ir(111) ($D_{ij}=1.0$~meV/Fe)~\cite{Dupe2014} and Co/Pt(111) ($D_{ij}=1.8$~meV/Co)~\cite{Dupe2014,Yang2015b}.

In the case of Co/Ru(0001), both the MAE and the DMI are in the range or smaller than 0.1 meV/atom. In that energy range, the dipole-dipole interaction should be considered in the stability of the different magnetic phase. We have evaluated the dipole-dipole contribution to the total energy for a structural domain of size $270 \times 11.7$~nm$^2$~(see Supplementary S7) and we find that the spin spiral has a negligible contribution to the dipole-dipole energy density. In addition, the difference of energy density between an in-plane and an out-of-plane homogenously
magnetized domain is $72 \, \mu$eV/Co and $70 \, \mu$eV/Co when boundary conditions are periodic and open, respectively. The obtained low values of the  DMI and MAE are consistent with the experimental results, even though they are at the limit of precision of the calculations. Within these values, 
 a small increase of the DMI, stabilizes the spin-spiral ground state as shown in Fig.~\ref{fig_T1}a. When $D_{ij}=0.3$~meV/Co, the ground state is a spin spiral for $\kappa_{\mathrm{FM}}=15 \, \mu$eV/Co. To obtain spin-spirals as the ground state, a slightly larger value of $D_{ij}$ is required or a smaller MAE. In all these case, isolated skyrmions would remain metastable. To further control our coefficients, we performed Monte Carlo (MC) simulations (see Supplementary Fig. S5) to obtain the critical temperature $T_{c}$. We found $T_c=150$~K in relative good agreement with experimental findings $T_c$ of $150$~K~\cite{SPLEEM}.

At surfaces and interfaces, the model of Levy and Fert~\cite{Fert1980} approximates $\mathbf{D}_{ij}$ as:

\begin{equation}
	\mathbf{D}_{ij}=V \left( \lambda_D \right) \left( \mathbf{r}_i \times \mathbf{r}_j \right),
	\label{Dij}
\end{equation}

\noindent where $\mathbf{r}_i$ and $\mathbf{r}_j$ are the position of two magnetic atoms (in Co) with respect to a non-magnetic atom (Ru in our case) and $\lambda_D$ is the strength of SOC of the $d$-electrons of the non-magnetic atom on the conduction electrons. Figure \ref{fig_T1}b shows the total contribution (red line), the Co contribution (black line) and the Ru contribution (green line) to the SOC. In the case of Co/Ru(0001), the DMI is induced by the SOC of Co and not of the Ru substrate in contrast to previous works \cite{,Dupe2016b,Dupe2014,Yang2015b,Simon2014}. In that respect, Co/Ru(0001) is the first ultra-thin film where the spin spiral ground state is induced by the SOC of the magnetic monolayer.

\begin{figure}
	\begin{center}
		\includegraphics[width=7.5cm]{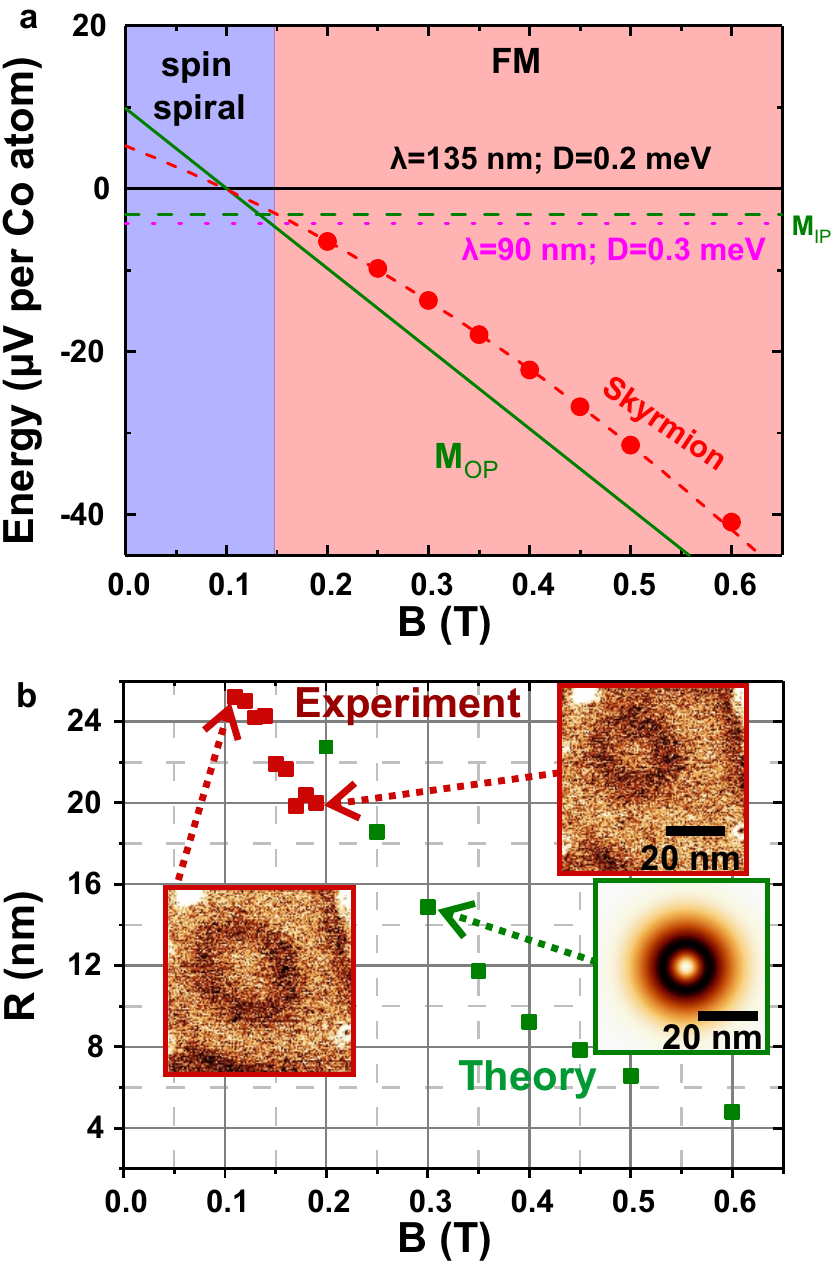}
	\end{center}
	\caption{\label{fig_4}
		{\bf Magnetic phase diagram of Co/Ru(0001).}
		(a) Energy of two spin spirals ($D=0.2$~meV in black and $D=0.3$~meV in magenta), isolated skyrmions and FM states as a function of the magnetic field $B$ obtained via spin dynamics simulations. The origin of energy is set to the left rotating spin spiral energy minimum obtained via DFT calculations (black line). The period the spin spiral is found to be $\lambda=135$~nm. The FM state is represented as a continuous green line when $\mathbf{M}$ is aligned perpendicular to the surface ($\mathbf{u}_z$~direction) and as a dashed line when $\mathbf{M}$ is in-plane. In the case of $D=0.2$~meV/Co, the ground is FM as expected from DFT calculations. For $D=0.3$~meV, the ground state is a spin spiral of period $\lambda=90$~nm (magenta dotted line). The energy of isolated skyrmions is shown as dashed red line. When $B\in[150,200]$~mT, the energy difference between the different magnetic states is as small as $2\,\mu$eV/Co. Isolated skyrmions are metastable in this region. (b) Dependency of the skyrmion radius with the magnetic field - Green dots correspond to to radius extracted from Monte Carlo simulations - Red dots are extracted from experiments realized with a bare tungsten tip - inset red frame: two example of the experimental $dI/dU$ map used to extract this profile showing a single skyrmion through the TAMR effect - the left one was recorded at \unit{120}{\milli\tesla} - the right one at \unit{190}{\milli\tesla} (\unit{60}{\nano\meter}$\times$\unit{60}{\nano\meter}, $I_{t}$=\unit{1}{\nano\ampere}, $U$=\unit{-220}{\milli\volt}, $\Delta{U}^{rms}$=\unit{50}{\milli\volt}) - inset green frame: Theoretical $dI/dU$ TAMR contrast of an isolated skyrmion at \unit{300}{\milli\tesla} (\unit{55}{\nano\meter}$\times$\unit{55}{\nano\meter})}
\end{figure}

In order to study the stability of the different phases, we minimized their total energies via spin dynamics simulations. We solve the Landau-Lifschitz-Gilbert equation using the extended Heisenberg model parametrized from our DFT calculation. Figure~\ref{fig_4}a shows the energy of the different phases as a function of the applied magnetic field with respect to the energy obtained for the spin spiral with $D_{ij}=0.2$~meV/Co. 
As discussed above, when $D_{ij}=0.2$~meV/Co, the ground state of the system at $B=0$~T is FM with in-plane magnetization (green dashed line). $D=0.3$~meV/Co is enough to stabilize a spin-spiral ground state for $B$ lower than $\approx$150 mT (pink dotted line). When $B$ further increases, the FM state alignes to the perpendicular magnetic field. The out of plane the magnetized FM state becomes more stable (green solid line). In the entire range of applied field, we have calculated the energy of isolated skyrmions (shown as red dashed lines). Isolated skyrmions are always metastable with respect to both the FM and the spin spiral states. The energy density differences are very small. For example, at $B=130$~mT, the energy difference between the isolated skyrmion state and the FM state is of the order of 0.5 $\mu$eV/Co.

Finally, we have studied the dependence of the skyrmion size on a small perpendicular magnetic field. As the skyrmions can be stabilized at rather low fields, any stray field of the tip would modify the skyrmion structure or laterally move it during scanning. To avoid this, a bare tungsten tip was used to image the skyrmion with the TAMR at -\unit{220}{\milli\volt}. We varied the perpendicular field within the experimentally accessible range between  110 and \unit{190}{\milli\tesla} in the direction antiparallel to the core of the skyrmion, i.e. the field is expected to compress the skyrmion. Two of the $dI/dU$ maps recorded are displayed in the inset of Fig. \ref{fig_4}b. From 120 to \unit{190}{\milli\tesla}, the diameter of the dark ring, i.e. the in-plane oriented section, is reduced by about \unit{5}{\nano\meter}. Here, the radius was determined by fitting the $dI/dU$ data to a 2D radial cosine function \cite{Moreau-Luchaire2016} (see Supplementary S6). 
Fig. \ref{fig_4}b also contains the theoretical radius obtained from the simulations for magnetic fields above 200 mT, where skyrmions showed stable radii. The experimental (red points) and theoretical radii (green points) agree well within few nm precision. This further confirms that the parameters of the calculations (MAE, DMI, Heisenberg exchange and dipolar energies) agree well with the experiments.

In conclusion, we demonstrated that even in systems with a weak DMI, skyrmions can be stabilized at low magnetic fields in case the MAE is sufficiently low. This approach allows one to manipulate skyrmions with fields suitable to application. In addition, these skyrmions are of tens of nanometers size facilitating their detection in lithographically structured samples and at the same time lifting the materials constraints on the interfacial layers (heavy 5d elements). We predict that this approach will quickly find utilization in proof-of-principle devices.

     
{\bf Methods}

Experiments: Sample and tips were prepared under ultra-high vacuum (UHV) at a base pressure of 4$\times$10$^{-11}$ mbar. Unpolarized STM tips were prepared from a W wire and were cleaned {\it in-situ} by flashing above \unit{2800}{\celsius}. Spin-polarized tips were prepared by depositing a Cr thin film onto the tip followed by a gentle annealing \cite{Wiesendanger}. The Ru(0001) single crystal was cleaned by cycles of annealing in oxygen at \unit{1000}{\celsius} followed by flashing to \unit{1500}{\celsius}. Once the substrate was depleted from bulk carbon impurities, cycles of argon-ion sputtering and annealing to \unit{1500}{\celsius} were performed to obtain atomically flat and clean surfaces. The Co film was deposited from an e-beam evaporator onto the clean Ru surface with a deposition rate of 0.3 ML per minute. Tips and samples were directly transferred to the STM under UHV. STM measurements were performed at \unit{4.2}{\kelvin} with a home-built microscope.

Density Functional Theory: For the DFT study the FLAPW basis as implemented in the FLEUR ab-initio package was used, which  accurately describes the ultra-thin film geometry by considering different basis function for the vacuum, the atomic muffin tin (MT) and the interstitial region. Within this framework, we have relaxed a symmetric Co/Ru(0001) slab composed of two Co monolayers separated by five layers of Ru. We have used a mixed LDA/GGA exchange and correlation functional that treats the MT of Ru in LDA and the MT of Co in GGA. In detail, we have used the Vosko LDA~\cite{VWN} and the GGA-PBE \cite{Perdew1992b}. In order to ensure a good convergence, we have used a cut-off of the plane waves basis set (K$_{\mathrm{max}}$) of $4.0$~bohr$^{-1}$ and $110$~k$-$points in $1/12^{\mathrm{th}}$ of the first Brillouin zone (BZ) of the hexagonal unit cell. The muffin tin radius of Co and Ru are $2.27$ bohr and $2.4$ bohr, respectively.
After the structural relaxation was performed, the spin-spiral energies were calculated in a unit cell containing one Co atom and 5 Ru atoms. The dispersion curves were calculated via the generalized Bloch theorem. The SOC contribution was calculated via first order perturbation theory~\cite{Heide20092678}. The magnetic force theorem~\cite{PhysRevB.39.86} was used to calculate the 54 nm and the 27 nm spin spiral with and without SOC contribution. We have used a $300 \times 300$~k-points mesh and K$_{\mathrm{max}}$ was set to $4.3$~bohr$^{-1}$ for both calculations.
We have calculated the MAE as the energy difference between several magnetic configurations in a supercell containing one Co on 9 Ru layers.  We have considered three different cases: An out-of-plane easy axis, an in-plane easy axis along the $\bar{\Gamma}-\bar{K}$ and along the $\bar{\Gamma}-\bar{M}$ direction. All magnetic configurations were converged self-consistently. We have used K$_{\mathrm{max}}=4.3$~bohr$^{-1}$, $44 \times 44$,  $63 \times 63$ and a $83 \times 83$~k-points mesh in the full BZ. The calculations give a MAE of $15$~$\mu$eV/Co, $40$~$\mu$eV/Co and $50$~$\mu$eV/Co, respectively. We have used $15$~$\mu$eV/Co for the Monte Carlo and the spin dynamics simulations. Higher values of MAE would not change the good qualitative agreement between theory and experiments.

Spin dynamics simulations: We have mapped our DFT calculations on the magnetic Hamiltonian:

\begin{align}
	H = -\sum_{ij} J_{\mathrm{eff}} \mathbf{M}_i \cdot \mathbf{M}_j - \sum_{ij} \mathbf{D}_{ij} \cdot \left(  \mathbf{M}_i \times \mathbf{M}_j \right) \\
	& \hspace*{-6.25cm} + \sum_{i} K \left( M^z_i\right)^2 + \sum_{i} \mathbf{B} \cdot \mathbf{M}_i ,
\end{align}

\noindent where $J_{\mathrm{eff}}$ is the magnetic exchange interaction close to the FM ground state, $\mathbf{D}_{ij}$ is the Dzyaloshinskii-Moriya interaction, $K$ is the uniaxial anisotropy vector and $\mathbf{B}$ is the external magnetic field.
The profile and energy stability of the isolated skyrmions were relaxed via spin dynamics using the Landau-Lifschitz-Gilbert equation (LLG):

\begin{equation*}
	\hbar \frac{ d \mathbf{M} }{dt}= \mathbf{M} \times \mathbf{B_{\mathrm{eff}}} - \alpha \mathbf{M} \times \left(\mathbf{M} \times \mathbf{B_{\mathrm{eff}}} \right) ,
\end{equation*}

\noindent where $\hbar$ is the Planck constant and $\mathbf{B}_{\mathrm{eff}}=-d H/d \mathbf{M}$ is the effective field created by the neighboring magnetic moments. We have integrated the LLG equations with a Heun integrator. The maximal torque $\parallel \mu_0\mathbf{B_{\mathrm{eff}}} \times \mathbf{M} \parallel$ was converge down to $2.10^{-6}$~eV.

{\bf Acknowledgements} \\
The authors would like to thank Stefan Heinze, Arthur Ernst and Khalil Zakeri for useful discussion. B.D. acknowledges HLRN Mogon for computation time. M.H. acknowledges funding by the European Commission (Grant ATOMS FP7/2007-2013-62260), W.W. funding by the Deutsche Forschungsgemeinschaft (grant Wu349/15-1), M.D.M. and R.L. funding by the Brazilian agencies CAPES, CNPq and FAPEMIG. B.D., M.B. and J.S. acknowledge the Alexander von Humboldt Foundation, the Deutsche Forschungsemeinschaft (grant DU1489/2-1), the Graduate School Materials Mainz, the ERC Synergy Grant SC2 (No. 610115), the Transregional Collaborative Research Center (SFB/TRR) 173 SPIN+X, and the Grant Agency of the Czech Republic grant no. 14-37427G. 

{\bf Author contributions} \\
W.W., M.D.M. and M.H. conceived the study. M.H., R.L, M.D.M. and L.G. performed the STM experiments. M.H. and T.B. analyzed the experimental data. B.D. performed and analyzed the DFT calculations. B.D. and M.B. performed and analyzed the Monte Carlo simulations. M.H., B.D., W.W. and J.S. wrote the manuscript. All authors discussed the data and reviewed the manuscript.  

\bibliography{citations}

\begin{thebibliography}{44}
\expandafter\ifx\csname natexlab\endcsname\relax\def\natexlab#1{#1}\fi

\bibitem[{Bogdanov \& Hubert(1994)}]{Bocdanov-1994aa}
Bogdanov, A.~N. \& Hubert, A.
\newblock The properties of isolated magnetic vortices.
\newblock \emph{Phys. stat. solid.} \textbf{186}, 527--543
\newblock  (1994).

\bibitem[{R{\"{o}}ssler \emph{et~al.}(2006)R{\"{o}}ssler, Bogdanov \&
  Pfleiderer}]{nature05056}
R{\"{o}}ssler, U.~K., Bogdanov, A.~N. \& Pfleiderer, C.
\newblock Spontaneous skyrmion ground states in magnetic metals.
\newblock \emph{Nature} \textbf{442}, 797--801
\newblock  (2006).

\bibitem[{Bogdanov \& Yablonskii(1989)}]{Bogdanov1989}
Bogdanov, A.~N. \& Yablonskii, D.
\newblock Thermodynamically stable "vortices" in magnetically ordered crystals.
  The mixed state of magnets.
\newblock \emph{Zh. Eksp. Teor. Fiz.} \textbf{95}, 178--182
\newblock  (1989).

\bibitem[{Jonietz \emph{et~al.}(2010)}]{Jonietz1648}
Jonietz, F. \emph{et~al.}
\newblock Spin transfer torques in MnSi at ultralow current densities.
\newblock \emph{Science} \textbf{330}, 1648--1651
\newblock  (2010).

\bibitem[{Fert \emph{et~al.}(2013)Fert, Cros \& Sampaio}]{nnano.2013.29}
Fert, A., Cros, V. \& Sampaio, J.
\newblock Skyrmions on the track.
\newblock \emph{Nat. Nanotechnol.} \textbf{8}, 152--156
\newblock  (2013).

\bibitem[{Sampaio \emph{et~al.}(2013)Sampaio, Cros, Rohart, Thiaville \&
  Fert}]{Sampaio2013}
Sampaio, J., Cros, V., Rohart, S., Thiaville, A. \& Fert, A.
\newblock Nucleation, stability and current-induced motion of isolated magnetic
  skyrmions in nanostructures.
\newblock \emph{Nat. Nanotechnol.} \textbf{8}, 839--844
\newblock  (2013).

\bibitem[{Dzyaloshinsky(1958)}]{Dzyaloshinsky1958}
Dzyaloshinsky, I.~E.
\newblock A thermodynamic theory of “weak” ferromagnetism of
  antiferromagnetics.
\newblock \emph{J. Phys. Chem. Solids} \textbf{4}, 241--255
\newblock  (1958).

\bibitem[{Dzyaloshinskii(1964)}]{Dzyaloshinskii1964}
Dzyaloshinskii, I.
\newblock Theory of helicoidal structures in antiferromagnets. I. Nonmetals.
\newblock \emph{Sov. Phys. JETP} \textbf{19}, 960--971
\newblock  (1964).

\bibitem[{Moriya(1960)}]{PhysRev.120.91}
Moriya, T.
\newblock Anisotropic superexchange interaction and weak ferromagnetism.
\newblock \emph{Phys. Rev.} \textbf{120}, 91--98
\newblock  (1960).

\bibitem[{M{\"{u}}hlbauer \emph{et~al.}(2009)}]{muhlbauer2009skyrmion}
M{\"{u}}hlbauer, S. \emph{et~al.}
\newblock Skyrmion lattice in a chiral magnet.
\newblock \emph{Science} \textbf{323}, 915--919
\newblock  (2009).

\bibitem[{Moreau-Luchaire \emph{et~al.}(2016)}]{Moreau-Luchaire2016}
Moreau-Luchaire, C. \emph{et~al.}
\newblock Additive interfacial chiral interaction in multilayers for
  stabilization of small individual skyrmions at room temperature.
\newblock \emph{Nat. Nanotechnol.} \textbf{11}, 444--448
\newblock  (2016).

\bibitem[{Dup{\'{e}} \emph{et~al.}(2016)Dup{\'{e}}, Bihlmayer, B{\"{o}}ttcher,
  Bl{\"{u}}gel \& Heinze}]{Dupe2016b}
Dup{\'{e}}, B., Bihlmayer, G., B{\"{o}}ttcher, M., Bl{\"{u}}gel, S. \& Heinze,
  S.
\newblock Engineering skyrmions in transition-metal multilayers for
  spintronics.
\newblock \emph{Nat. Commun.} \textbf{7}, 11779
\newblock  (2016).

\bibitem[{Heinze \emph{et~al.}(2011)}]{nphys2045}
Heinze, S. \emph{et~al.}
\newblock Spontaneous atomic-scale magnetic skyrmion lattice in two dimensions.
\newblock \emph{Nat. Phys.} \textbf{7}, 713--718
\newblock  (2011).

\bibitem[{Jiang \emph{et~al.}(2015)}]{Jiang2015a}
Jiang, W. \emph{et~al.}
\newblock Blowing magnetic skyrmion bubbles.
\newblock \emph{Science} \textbf{349}, 283--286
\newblock  (2015).

\bibitem[{{Hrabec} \emph{et~al.}(2017)}]{2016arXiv161100647H}
{Hrabec}, A. \emph{et~al.}
\newblock {Current-induced skyrmion generation and dynamics in symmetric
  bilayers}.
\newblock \emph{Nat. Commun.} \textbf{8}, 15765
\newblock  (2017).

\bibitem[{{Pollard} \emph{et~al.}(2017)}]{Pollard2017}
{Pollard}, S.~D. \emph{et~al.}
\newblock {Observation of stable Néel skyrmions in cobalt/palladium
  multilayers with Lorentz transmission electron microscopy}.
\newblock \emph{Nat. Commun.} \textbf{8}, 14761
\newblock  (2017).

\bibitem[{Hsu \emph{et~al.}(2017)}]{hsu2017}
Hsu, P. \emph{et~al.}
\newblock {Electric-field-driven switching of individual magnetic skyrmions}.
\newblock \emph{Nat. Nano.} \textbf{12}, 123--126
\newblock  (2017).

\bibitem[{Romming \emph{et~al.}(2013)}]{Romming2013}
Romming, N. \emph{et~al.}
\newblock Writing and deleting single magnetic skyrmions.
\newblock \emph{Science} \textbf{341}, 636--639
\newblock  (2013).

\bibitem[{Nandy \emph{et~al.}(2016)Nandy, Kiselev \& Bl{\"{u}}gel}]{Nandy2016}
Nandy, A.~K., Kiselev, N.~S. \& Bl{\"{u}}gel, S.
\newblock Interlayer Exchange Coupling: A General Scheme Turning Chiral Magnets
  into Magnetic Multilayers Carrying Atomic-Scale Skyrmions.
\newblock \emph{Phys. Rev. Lett.} \textbf{116}, 177202
\newblock  (2016).

\bibitem[{Dup{\'{e}} \emph{et~al.}(2014)Dup{\'{e}}, Hoffmann, Paillard \&
  Heinze}]{Dupe2014}
Dup{\'{e}}, B., Hoffmann, M., Paillard, C. \& Heinze, S.
\newblock Tailoring magnetic skyrmions in ultra-thin transition metal films.
\newblock \emph{Nat. Commun.} \textbf{5}, 4030
\newblock  (2014).

\bibitem[{Simon \emph{et~al.}(2014)Simon, Palot{\'{a}}s, R{\'{o}}zsa, Udvardi
  \& Szunyogh}]{Simon2014}
Simon, E., Palot{\'{a}}s, K., R{\'{o}}zsa, L., Udvardi, L. \& Szunyogh, L.
\newblock Formation of magnetic skyrmions with tunable properties in PdFe
  bilayer deposited on Ir(111).
\newblock \emph{Phys. Rev. B} \textbf{90}, 094410
\newblock  (2014).

\bibitem[{Leonov \emph{et~al.}(2016)}]{Leonov2015a}
Leonov, A. \emph{et~al.}
\newblock The properties of isolated chiral skyrmions in thin magnetic films.
\newblock \emph{New J. Phys.} \textbf{18}, 065003
\newblock  (2016).

\bibitem[{Rohart \& Thiaville(2013)}]{Rohart2013}
Rohart, S. \& Thiaville, A.
\newblock Skyrmion confinement in ultrathin film nanostructures in the presence
  of Dzyaloshinskii-Moriya interaction.
\newblock \emph{Phys. Rev. B} \textbf{88}, 184422
\newblock  (2013).

\bibitem[{Gabaly \emph{et~al.}(2007)}]{growth}
Gabaly, F.~E. \emph{et~al.}
\newblock Structure and morphology of ultrathin Co/Ru(0001) films.
\newblock \emph{New Journal of Physics} \textbf{9}, 80
\newblock  (2007).

\bibitem[{Baltz \emph{et~al.}(2007)Baltz, Marty, Rodmacq \&
  Dieny}]{PhysRevB.75.014406}
Baltz, V., Marty, A., Rodmacq, B. \& Dieny, B.
\newblock Magnetic domain replication in interacting bilayers with out-of-plane
  anisotropy: Application to $\mathrm{Co}∕\mathrm{Pt}$ multilayers.
\newblock \emph{Phys. Rev. B} \textbf{75}, 014406
\newblock  (2007).

\bibitem[{{Phark} \emph{et~al.}(2014)}]{sanders}
{Phark}, S.~H. \emph{et~al.}
\newblock {Reduced-dimensionality-induced helimagnetism in iron nanoislands}.
\newblock \emph{Nat. Commun.} \textbf{5}, 5183
\newblock  (2014).

\bibitem[{Bode \emph{et~al.}(2007)}]{BodeNature}
Bode, M. \emph{et~al.}
\newblock Chiral magnetic order at surfaces driven by inversion asymmetry.
\newblock \emph{Nature} \textbf{447}, 190--193
\newblock  (2007).

\bibitem[{Bode \emph{et~al.}(2002)}]{bode}
Bode, M. \emph{et~al.}
\newblock Magnetization-Direction-Dependent Local Electronic Structure Probed
  by Scanning Tunneling Spectroscopy.
\newblock \emph{Phys. Rev. Lett.} \textbf{89}, 237205
\newblock  (2002).

\bibitem[{Gould \emph{et~al.}(2004)}]{TAMR}
Gould, C. \emph{et~al.}
\newblock Tunneling Anisotropic Magnetoresistance: A Spin-Valve-Like Tunnel
  Magnetoresistance Using a Single Magnetic Layer.
\newblock \emph{Phys. Rev. Lett.} \textbf{93}, 117203
\newblock  (2004).

\bibitem[{von Bergmann \emph{et~al.}(2012)}]{VanBergmann}
von Bergmann, K. \emph{et~al.}
\newblock Tunneling anisotropic magnetoresistance on the atomic scale.
\newblock \emph{Phys. Rev. B} \textbf{86}, 134422
\newblock  (2012).

\bibitem[{Ferriani \emph{et~al.}(2008)}]{bode2}
Ferriani, P. \emph{et~al.}
\newblock Atomic-Scale Spin Spiral with a Unique Rotational Sense: Mn Monolayer
  on W(001).
\newblock \emph{Phys. Rev. Lett.} \textbf{101}, 027201
\newblock  (2008).

\bibitem[{Romming \emph{et~al.}(2015)Romming, Kubetzka, Hanneken, von Bergmann
  \& Wiesendanger}]{Romming2015}
Romming, N., Kubetzka, A., Hanneken, C., von Bergmann, K. \& Wiesendanger, R.
\newblock Field-Dependent Size and Shape of Single Magnetic Skyrmions.
\newblock \emph{Phys. Rev. Lett.} \textbf{114}, 177203
\newblock  (2015).

\bibitem[{Haze \emph{et~al.}(2017)Haze, Yoshida \&
  Hasegawa}]{PhysRevB.95.060415}
Haze, M., Yoshida, Y. \& Hasegawa, Y.
\newblock Role of the substrate in the formation of chiral magnetic structures
  driven by the interfacial Dzyaloshinskii-Moriya interaction.
\newblock \emph{Phys. Rev. B} \textbf{95}, 060415
\newblock  (2017).

\bibitem[{FLE(????)}]{FLEUR}
www.flapw.de
\newblock  (????).

\bibitem[{Hardrat \emph{et~al.}(2009)}]{PhysRevB.79.094411}
Hardrat, B. \emph{et~al.}
\newblock Complex magnetism of iron monolayers on hexagonal transition metal
  surfaces from first principles.
\newblock \emph{Phys. Rev. B} \textbf{79}, 094411
\newblock  (2009).

\bibitem[{Kurz \emph{et~al.}(2004)Kurz, F{\"{o}}rster, Nordstr{\"{o}}m,
  Bihlmayer \& Bl{\"{u}}gel}]{PhysRevB.69.024415}
Kurz, P., F{\"{o}}rster, F., Nordstr{\"{o}}m, L., Bihlmayer, G. \&
  Bl{\"{u}}gel, S.
\newblock Ab initio treatment of noncollinear magnets with the full-potential
  linearized augmented plane wave method.
\newblock \emph{Phys. Rev. B} \textbf{69}, 024415
\newblock  (2004).

\bibitem[{Bruno(1989)}]{PhysRevB.39.86}
Bruno, P.
\newblock Tight-binding approach to the orbital magnetic moment and
  magnetocrystalline anisotropy of transition-metal monolayers.
\newblock \emph{Phys. Rev. B} \textbf{39}, 865--868
\newblock  (1989).

\bibitem[{El~Gabaly \emph{et~al.}(2006)}]{SPLEEM}
El~Gabaly, F. \emph{et~al.}
\newblock Imaging Spin-Reorientation Transitions in Consecutive Atomic Co
  Layers on Ru(0001).
\newblock \emph{Phys. Rev. Lett.} \textbf{96}, 147202
\newblock  (2006).

\bibitem[{Yang \emph{et~al.}(2015)Yang, Thiaville, Rohart, Fert \&
  Chshiev}]{Yang2015b}
Yang, H., Thiaville, A., Rohart, S., Fert, A. \& Chshiev, M.
\newblock Anatomy of Dzyaloshinskii-Moriya Interaction at Co / Pt Interfaces.
\newblock \emph{Phys. Rev. Lett.} \textbf{115}, 267210
\newblock  (2015).

\bibitem[{Fert \& Levy(1980)}]{Fert1980}
Fert, A. \& Levy, P.~M.
\newblock Role of anisotropic exchange interactions in determining the
  properties of spin-glasses.
\newblock \emph{Phys. Rev. Lett.} \textbf{44}, 1538--1541
\newblock  (1980).

\bibitem[{Wiesendanger(2009)}]{Wiesendanger}
Wiesendanger, R.
\newblock Spin mapping at the nanoscale and atomic scale.
\newblock \emph{Rev. Mod. Phys.} \textbf{81}, 1495--1550
\newblock  (2009).

\bibitem[{Vosko \emph{et~al.}(1980)Vosko, Wilk \& Nusair}]{VWN}
Vosko, S.~H., Wilk, L. \& Nusair, M.
\newblock Accurate spin-dependent electron liquid correlation energies for
  local spin density calculations: a critical analysis.
\newblock \emph{Can. J. Phys.} \textbf{58}, 1200--1211
\newblock  (1980).

\bibitem[{Perdew \emph{et~al.}(1992)}]{Perdew1992b}
Perdew, J.~P. \emph{et~al.}
\newblock Atoms, molecules, solids, and surfaces: Applications of the
  generalized gradient approximation for exchange and correlation.
\newblock \emph{Phys. Rev. B} \textbf{46}, 6671--6687
\newblock  (1992).

\bibitem[{Heide \emph{et~al.}(2009)Heide, Bihlmayer \&
  Bl{\"{u}}gel}]{Heide20092678}
Heide, M., Bihlmayer, G. \& Bl{\"{u}}gel, S.
\newblock Describing Dzyaloshinskii-Moriya spirals from first principles.
\newblock \emph{Phys. B} \textbf{404}, 2678--2683
\newblock  (2009).

\end{thebibliography}


\begin{thebibliography}{8}
\expandafter\ifx\csname natexlab\endcsname\relax\def\natexlab#1{#1}\fi

\bibitem[{Gabaly \emph{et~al.}(2007)}]{growth}
Gabaly, F.~E. \emph{et~al.}
\newblock Structure and morphology of ultrathin Co/Ru(0001) films.
\newblock \emph{New Journal of Physics} \textbf{9}, 80
\newblock  (2007).

\bibitem[{Rodary \emph{et~al.}(2009)Rodary, Wedekind, Oka, Sander \&
  Kirschner}]{sander}
Rodary, G., Wedekind, S., Oka, H., Sander, D. \& Kirschner, J.
\newblock Characterization of tips for spin-polarized scanning tunneling
  microscopy.
\newblock \emph{Applied Physics Letters} \textbf{95}, 152513
\newblock  (2009).

\bibitem[{Bode \emph{et~al.}(2002)}]{bode}
Bode, M. \emph{et~al.}
\newblock Magnetization-Direction-Dependent Local Electronic Structure Probed
  by Scanning Tunneling Spectroscopy.
\newblock \emph{Phys. Rev. Lett.} \textbf{89}, 237205
\newblock  (2002).

\bibitem[{von Bergmann \emph{et~al.}(2012)}]{vanbergman}
von Bergmann, K. \emph{et~al.}
\newblock Tunneling anisotropic magnetoresistance on the atomic scale.
\newblock \emph{Phys. Rev. B} \textbf{86}, 134422
\newblock  (2012).

\bibitem[{Gould \emph{et~al.}(2004)}]{TAMR}
Gould, C. \emph{et~al.}
\newblock Tunneling Anisotropic Magnetoresistance: A Spin-Valve-Like Tunnel
  Magnetoresistance Using a Single Magnetic Layer.
\newblock \emph{Phys. Rev. Lett.} \textbf{93}, 117203
\newblock  (2004).

\bibitem[{Jonietz \emph{et~al.}(2010)}]{STT}
Jonietz, F. \emph{et~al.}
\newblock Spin Transfer Torques in MnSi at Ultralow Current Densities.
\newblock \emph{Science} \textbf{330}, 1648--1651
\newblock  (2010).

\bibitem[{El~Gabaly \emph{et~al.}(2006)}]{SPLEEM}
El~Gabaly, F. \emph{et~al.}
\newblock Imaging Spin-Reorientation Transitions in Consecutive Atomic Co
  Layers on Ru(0001).
\newblock \emph{Phys. Rev. Lett.} \textbf{96}, 147202
\newblock  (2006).

\bibitem[{Bogdanov \& Hubert(1994)}]{Bocdanov-1994aa}
Bogdanov, A.~N. \& Hubert, A.
\newblock The properties of isolated magnetic vortices.
\newblock \emph{Phys. status solidi} \textbf{186}, 527--543
\newblock  (1994).

\end{thebibliography}
\bibliographystyle{nature}
\newpage

\newpage

\end{document}


\newcommand{\By}{$\times$}
\newcommand{\SqrtBy}[2]{$\sqrt{#1}$\kern0.2ex$\times$\kern-0.2ex$\sqrt{#2}$}
\newcommand{\Degree}{$^\circ$}
\newcommand{\DegreeC}{$^\circ$C}
\newcommand{\Ohmcm}{$\Omega\cdot$cm}

\title{Stabilizing isolated skyrmions at low magnetic fields exploiting vanishing magnetic anisotropy} 

\author{Marie Herv\'e$^{1}$, Bertrand Dup\'e$^{2,3}$, Rafael Lopes$^{4}$, Marie B\"ottcher$^{2}$, Maximiliano D. Martins$^4$,   Timofey Balashov$^{1}$, Lukas Gerhard$^{5}$, Jairo Sinova$^{2,6}$, Wulf Wulfhekel$^{1,5}$ \\
	$^1$Physikalisches Institut, Karlsruhe Institute of Technology, 76131 Karlsruhe, Germany \\
	$^2$Institut f\"ur Physik, Johannes Gutenberg Universit\"at Mainz, D-55099 Mainz, Germany \\
	$^3$Institute of Theoretical Physics and Astrophysics, University of Kiel, 24098 Kiel, Germany \\
	$^4$Centro de Desenvolvimento da Tecnologia Nuclear, 31270-901, Belo Horizonte, Brazil \\
	$^5$Institute of Nanotechnology, Karlsruhe Institute of Technology,  76128 Karlsruhe, Germany \\
	$^6$Institute of Physics, Academy of Sciences of the Czech Republic, Cukrovarnick\'{a} 10, 162 53 Praha 6 Czech Republic \\}
{\bf Supplementary material} 
 
 \maketitle
{\bf S1 Interdiffusion between Co and Ru} 

\begin{figure}[h]
	\begin{center}
		\includegraphics[width=7cm]{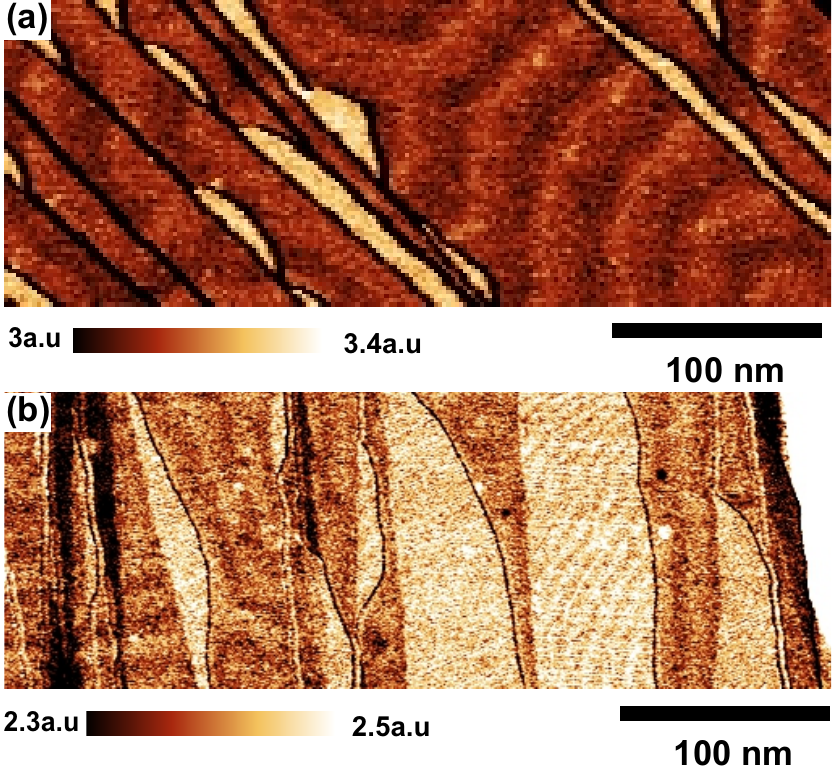}
		\caption{ \label{fig_S1} {\bf Alloy islands.}
			$dI/dU$ maps of 1.1 Co ML deposited at \unit{250}{\celsius} (a) and \unit{350}{\celsius} (b). The bright islands decorating the lower step edges are ascribed to an alloy of Co and Ru. The density of these islands increased with deposition temperature ($I$=\unit{1}{\nano\ampere}, $\Delta{U}_{rms}$=\unit{40}{\milli\volt}, (a) $U$=\unit{-400}{\milli\volt}, (b)  $U$=\unit{-290}{\milli\volt}). The pure Co areas show the spin spiral contrast probed with a spin-polarized tip in (a) and a non-spin polarized tip in (b).}
		
	\end{center}
\end{figure}

As mentioned in previous studies \cite{growth}, interdiffusion between Co and Ru occurs at deposition temperatures above \unit{250}{\celsius}. The STM topography Fig. 1a in the main text shows that for 1.1 Co ML deposited at $\approx$  \unit{300}{\celsius}. the lower step edges of Ru are decorated by islands appearing brighter (higher $dI/dU$) than the Co wetting layer in the $dI/dU$ map. Additional investigations of the density of these islands for different deposition temperatures are displayed in Fig. \ref{fig_S1}. 1.1 Co ML was deposited at \unit{250}{\celsius} (Fig. \ref{fig_S1}a) and \unit{350}{\celsius} (Fig. \ref{fig_S1}b). Increasing the deposition temperature from \unit{250}{\celsius} to \unit{350}{\celsius} leads to an increase of the density of these islands. We interpreted the nature of these islands as an alloying between Co and Ru. Detailed informations will be presented elsewhere.          

{\bf S2 Magnetic field dependency of the spin polarized contrast}

\begin{figure}[h]
	\begin{center}
		\includegraphics[width=12cm]{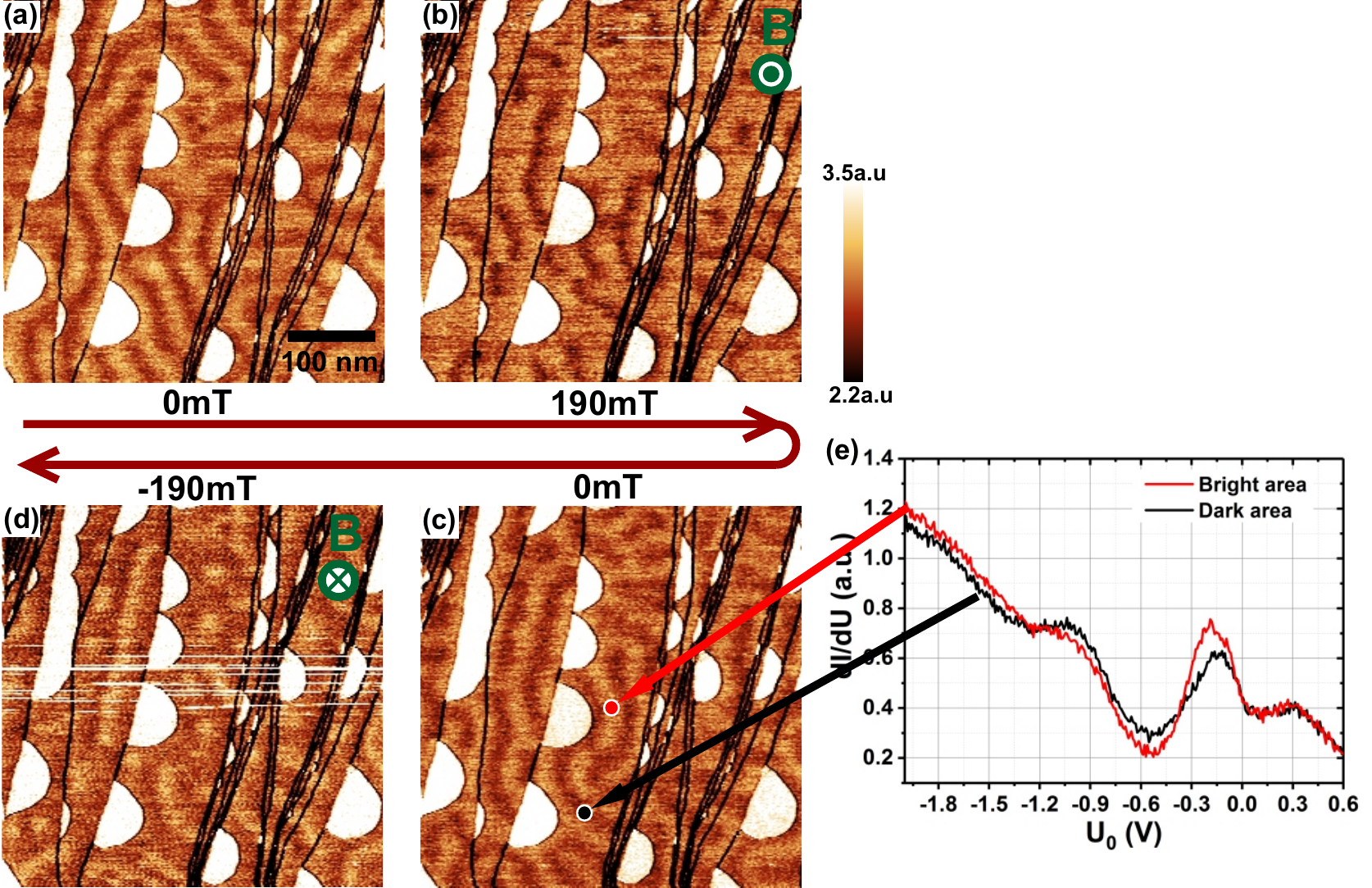}
	\end{center}
	\caption{\label{fig_S2}
		{\bf Magnetic field dependency of the stripe pattern.}
		(a) - (d) $dI/dU$ maps of the same area recorded consecutively with an out-of-plane spin-polarized tip at an out-of-plane magnetic field as indicated (400$\times$\unit{400}{\nano\meter}${}^{2}$, $I$=\unit{1}{\nano\ampere}, $U$=\unit{-500}{\milli\volt}, $\Delta{U}_{rms}$=\unit{40}{\milli\volt}). (e) $dI/dU$ spectra recorded in the middle of a bright stripe where the local magnetization is parallel to the tip spin polarization (red curve) and in the middle of a dark stripe where the local magnetization is antiparallel to the tip spin polarization (dark curve) ($I$=\unit{1}{\nano\ampere}, $U$=\unit{-1.5}{\volt}, $\Delta{U}_{rms}$=\unit{20}{\milli\volt}).}
\end{figure}

The periodic stripe pattern shown in Fig. 1c in the main text can be modified by the application of an out-of-plane magnetic field.  Moderate positive magnetic fields ($B$=\unit{+190}{\milli\tesla}, cf. Fig. \ref{fig_S2}b) lead to an expansion of the bright areas, while negative magnetic fields ($B$=\unit{-190}{\milli\tesla}), cf. Fig. \ref{fig_S2}d) lead to the expansion of the dark areas. Decreasing $B$ from \unit{+190}{\milli\tesla} to \unit{0}{\milli\tesla} regenerates the balance between dark and bright areas (cf. Fig. \ref{fig_S2}c). This indicates that the observed bright and dark contrast areas are due to a local out-of-plane orientation of magnetization. Note that for the small field used here, the spin-polarization direction of the tip is not affected \cite{sander}.

The voltage dependency of the TMR effect presented in Fig. 2a in the main text was obtained from the two curves presented in Fig. \ref{fig_S2}e. $dI/dU$ spectra were recorded by stabilizing the tip at -1.5 V, disabling the feedback and ramping the voltage with the tip located in the middle of a bright stripe (red curve) and in the middle of a dark stripe (dark curve). The relative variation of the $dI/dU$ signal between the spectra measured on the dark and the bright stripe was converted to the voltage dependency of the TMR (Fig. 2a - main text) as:                     
\begin{equation}
TMR=\frac{\frac{d I}{ d U}_{dark}-\frac{d I}{ d U}_{bright}}{\frac{d I}{ d U}_{dark}}.
\end{equation}

\newpage
{\bf S3 Magnetic field dependency of the TAMR contrast}

\begin{figure}[h]
	\begin{center}
		\includegraphics[width=10cm]{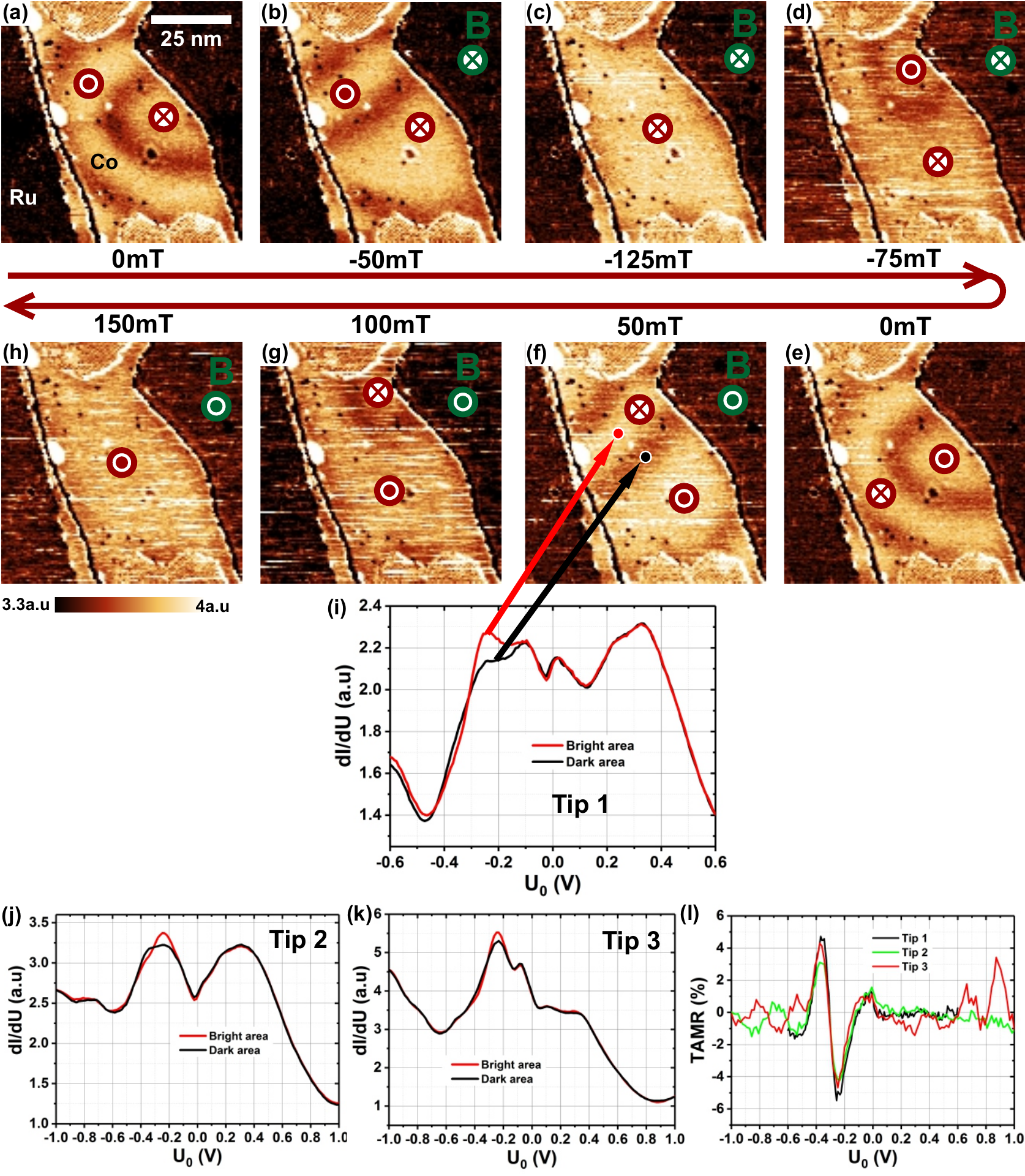}
	\end{center}
	\caption{\label{fig_S3}
		{\bf Magnetic field dependency of the TAMR contrast.}
		(a) -  (h) $dI/dU$ maps recorded at out-of-plane magnetic fields as indicated. The tip was unpolarized and the spin spiral was probed through TAMR ($I$=\unit{1}{\nano\ampere}, $U$=\unit{-250}{\milli\volt}, $\Delta{U}_{rms}$=\unit{30}{\milli\volt}). (i) - (k) $dI/dU$ spectra recorded on a bright stripe (red curve - local magnetization out-of-plane) and on a dark stripe (dark curve and local magnetization in-plane) with three different tips ((i): $I$=\unit{10}{\nano\ampere}, $U$=\unit{600}{\milli\volt}, $\Delta{U}_{rms}$=\unit{10}{\milli\volt} - (j) and (k): $I$=\unit{10}{\nano\ampere}, $U$=\unit{1000}{\milli\volt}, $\Delta{U}_{rms}$=\unit{20}{\milli\volt}). (l) Voltage dependency of the TAMR for the three different tips.}
\end{figure}

The evolution of the periodic stripe pattern under an out-of-plane magnetic field was recorded with non-spin polarized tip as well. Fig.  \ref{fig_S3}a shows a $dI/dU$ map recorded at a bias voltage of \unit{-250}{\milli\volt} on an isolated Co island. The darkest zone in this image  corresponds to the bare Ru substrate. The contrast inside the island shows the periodic stripe pattern. Fig.\ref{fig_S3}a to h display half of a magnetization cycle in this isolated Co island. For an out-of-plane magnetic field of \unit{-125}{\milli\tesla} (Fig. \ref{fig_S3}c) and \unit{+150}{\milli\tesla} (Fig. \ref{fig_S3}h), the magnetization is almost saturated and is aligned parallel with the magnetic field. For both positive and negative magnetic field, an expansion of the bright area is observed. The maximum of the $dI/dU$ signal (bright contrast) corresponds to an out-of-plane magnetization direction (either up or down). An in-plane magnetization gives a minimum of the $dI/dU$ signal (dark contrast). The observed contrast is interpreted as originating from an out-of-plane TAMR effect where the $dI/dU$ varies as a $cos^2$ with the angle formed between the magnetization and an out-of-plane quantization axis \cite{bode, vanbergman, TAMR}. Note that in the saturated magnetic state (Fig. \ref{fig_S3}c and h), dark contrast patches still remain on the island edges. At these locations, the magnetization is twisted. While decreasing the magnetic field toward \unit{0}{\milli\tesla} (Fig. \ref{fig_S3}d and e), the spin spiral state is regenerated.

The voltage dependency of the TAMR effect presented in Fig. 2a in the main text was obtained from the two curves presented in Fig. \ref{fig_S3}i. $dI/dU$ spectra were recorded with the tip located in the middle of a bright stripe (red curve - out-of-plane magnetization) and in the middle of a dark stripe (dark curve - in-plane magnetization). The relative variation of the $dI/dU$ signal is converted to the voltage dependency of the TAMR effect (Fig. 2a - main text)  as:                     
\begin{equation}
TAMR=\frac{\frac{d I}{ d U}_{in-plane}-\frac{d I}{ d U}_{out-of-plane}}{\frac{d I}{ d U}_{in-plane}}.
\end{equation}
           
As explained in the main text, an inversion of the TAMR occurs at \unit{-300}{\milli\volt}. Fig. \ref{fig_S3} shows the $dI/dU$ maps at a bias voltage of \unit{-250}{\milli\volt}. In that case, the TAMR is negative, a maximum of the $dI/dU$ signal corresponds to an out-of-plane magnetization, a minimum of the $dI/dU$ amplitude coincide with an in-plane magnetization. In contrast, Fig. 2c (main text) shows the $dI/dU$ map recorded at a bias voltage of \unit{-400}{\milli\volt}. Then, the TAMR is positive and the magnetization pointing out-of-plane correspond to a minimum of the $dI/dU$ amplitude, while in-plane magnetization shows a maximum of the $dI/dU$ signal. 

Note that in scanning tunneling spectroscopy measurements, sample and tip LDOS are convoluted. By means of voltage pulses it was possible to change the tip termination and its LDOS. Fig. \ref{fig_S3}j and k display two other kind of $dI/dU$ spectra recorded with two other tip termination. The general shape of the $dI/dU$ spectra is strongly affected by the tip LDOS. However, the TAMR feature (see Fig. \ref{fig_S3}l) remains unchanged.  

\newpage     
{\bf S4 Metastable skyrmions at zero magnetic field}

\begin{figure}[h]
	\begin{center}
		\includegraphics[width=12cm]{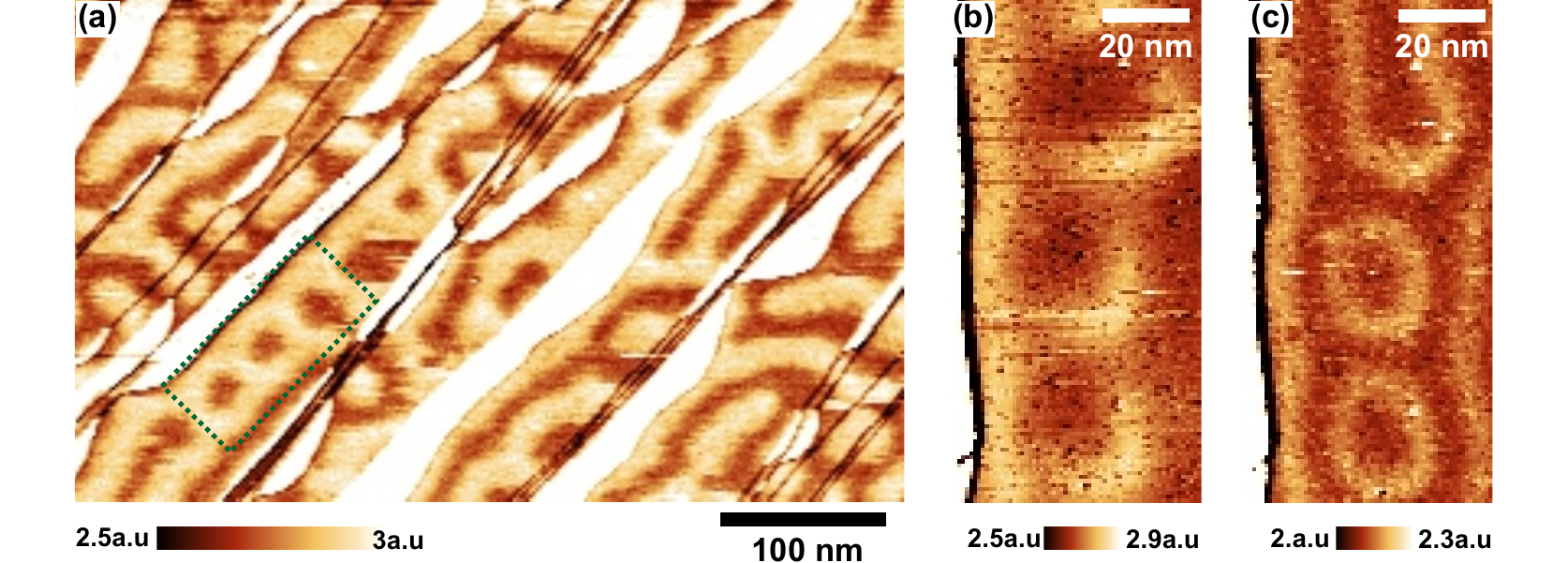}
	\end{center}
	\caption{\label{fig_S4}
		{\bf Metastable skyrmions at zero magnetic field.}
		$dI/dU$ maps showing remaining skyrmions after ramping down the magnetic field from \unit{-150}{\milli\tesla} to \unit{0}{\milli\tesla}. (a) The tip was out-of-plane spin-polarized ($I$=\unit{1}{\nano\ampere}, $U$=\unit{-400}{\milli\volt}, $\Delta{U}_{rms}$=\unit{40}{\milli\volt}). (b) and (c) were recorded in the area marked in (a) after two consecutive tip changes. The tip was in-plane spin-polarized in (b) and unpolarized in (c) ($I$=\unit{10}{\nano\ampere}, $U$=\unit{-400}{\milli\volt}, $\Delta{U}_{rms}$=\unit{40}{\milli\volt}).}
\end{figure}

As indicated in the main text, when ramping down the magnetic field to \unit{0}{\milli\tesla} some magnetic skyrmions remain stable. Fig. \ref{fig_S4}a displays a spin polarized $dI/dU$ map recorded after ramping down the magnetic field from \unit{-150}{\milli\tesla} to \unit{0}{\milli\tesla}. Tip was out-of-plane spin-polarized. On one of the narrow terraces (green dotted box) three skyrmions remain. Fig. \ref{fig_S4}b and c show two $dI/dU$ maps of this area recorded after two consecutive changes of the tip spin polarization. In (b), the tip was in-plane spin-polarized. The three skyrmions appear with the same orientation, which confirms that the spin structure is chiral. In (c) the tip was unpolarized and the skyrmions were imaged via the TAMR effect. Note that the top skyrmion of the image merged with a fourth skyrmion in (a). Indeed it is frequently observed that a spin polarized tip in this system interacts with the spin structure. The origin of this interaction is still unclear.

{\bf S5 Estimation of T$_{\mathrm{c}}$}

In order to compare the coefficients of our effective atomistic spin model with experiments, we performed Monte Carlo simulation to determine the critical temperature T$_{\mathrm{c}}$ at B=0 T. We have used a standard metropolis Monte Carlo method. To decrease the computational costs, we modeled a slab of Co/Ru(0001) using an asymmetric supercell composed of $200\times10$ magnetic moments along the $\bar{\Gamma}-\bar{M}$ direction. We have used $J_{\mathrm{eff}}=13.1$~meV/Co, $D_{ij}=-0.2$~meV/Co and $K=0.015$~meV/Co. The starting configuration is a ferromagnetic state where the magnetic moments are pointing along the long axis of the slab. We have computed 100 equally spaced temperatures within the range $[$1 K, 650 K$]$. For each temperature, the magnetic moments were first thermalized $2\times10^5$ MC steps. The energy, magnetization densities and the magnetic susceptibility were averaged over $5\times10^6$ MC steps, each of them separated by $2\times10^4$ autocorrelation steps.
\begin{figure}[ht]
	\begin{center}
		\includegraphics[width=8cm]{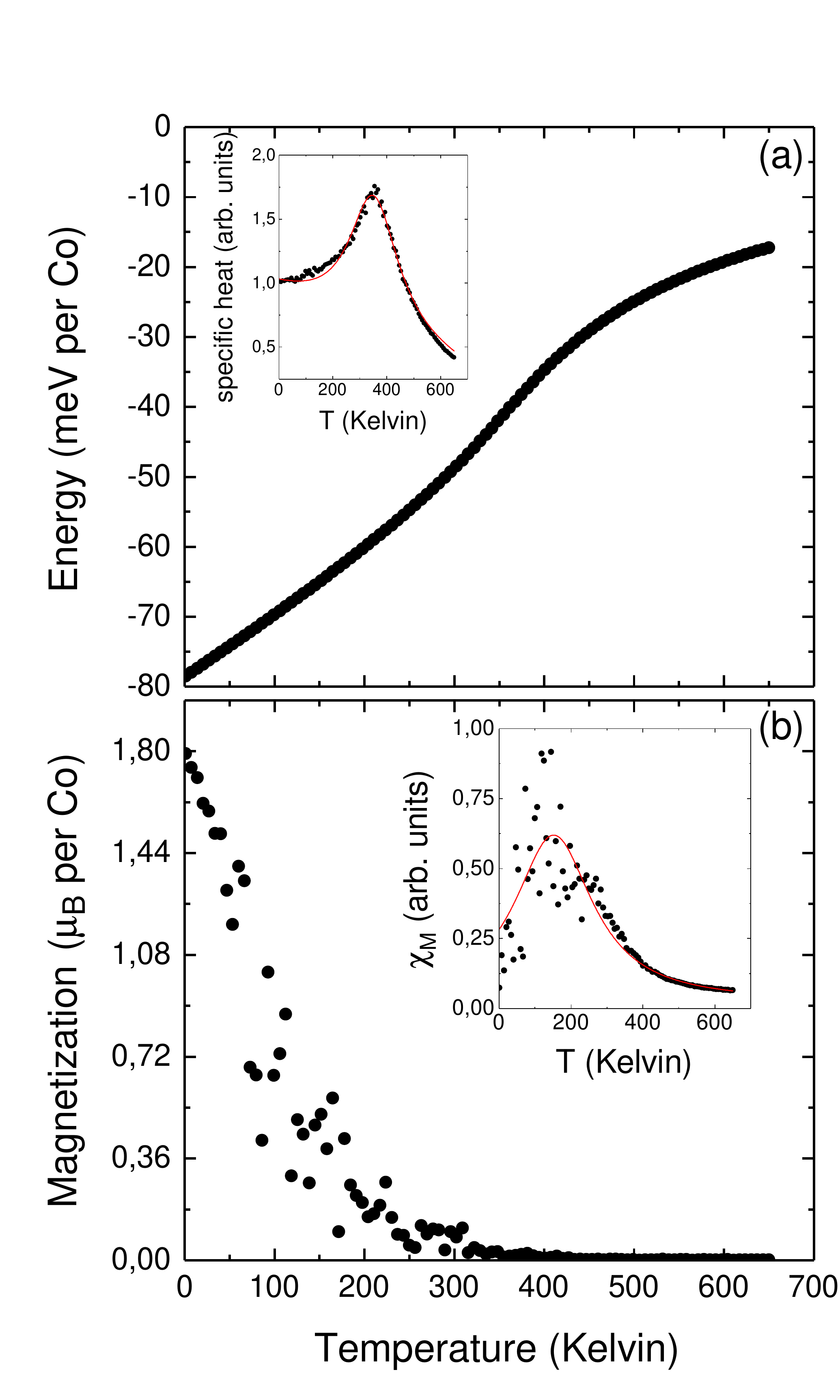}
	\end{center}
	\caption{\label{fig_S5}
		{\bf Determination of $T_{\mathrm{c}}$.}
		(a) Energy density as a function of temperature. The inset shows the specific heat as a function of temperature. The peak of specific heat corresponds to the phase transition between the ordered and disordered phases. (b) Magnetization density as a function of temperature. The magnetization disappears at $T\sim 200$ K. The inset shows the magnetic susceptibility and a lorentzian fit (red line). We extract from the fit $T^M_{\mathrm{c}}=152$ K.}
\end{figure}

In order to estimate the Curie temperature, we analyse the energy and the magnetization density. We compute the specific heat as:

\begin{eqnarray}
	C=\frac{ \left< E^2 \right> - \left< E \right>^2 }{k_{\mathrm{B}}T^2},
\end{eqnarray}

\noindent where $\left< E \right>$ is the average energy density, $k_{\mathrm{B}}$ is the Boltzmann constant and $T$ is the temperature. We have computed the magnetic susceptibility as:

\begin{eqnarray}
	\chi_{M}=\frac{ \left< M^2 \right> - \left< M \right>^2 }{k_{\mathrm{B}}T}.
\end{eqnarray}

We have modeled the energy and the magnetization density as arctangent function. We can then fit $C$ and $\chi_M$ with the Lorentzian function:

\begin{eqnarray}
	F(T)=F_0+\frac{2I}{\pi} \frac{\sigma}{4 \left( T - T_c \right)^2 + \sigma^2},
\end{eqnarray}

\noindent where $I$ is the intensity of the peak and $\sigma$ is the mean-height width. All the fits were carried out with a linear background. By fitting of the specific heat, we obtain $T_{\mathrm{c}}=350$~K. This critical temperature corresponds to the transition between the ordered phase to the disordered phase. Ref. \cite{SPLEEM} reports on the temperature at which the magnetization vanishes i.e. $T^{\mathrm{M}}_{\mathrm{c}}=170$~K. When the susceptibility $\chi_M$ is fitted with the Lorentzian curve, we found $T^{\mathrm{M}}_{\mathrm{c}}=152$~K, in good agreement with the experimental value.

{\bf S6 Determination of the skyrmion radii}

\begin{figure}[h]
	\begin{center}
		\includegraphics[width=8.5cm]{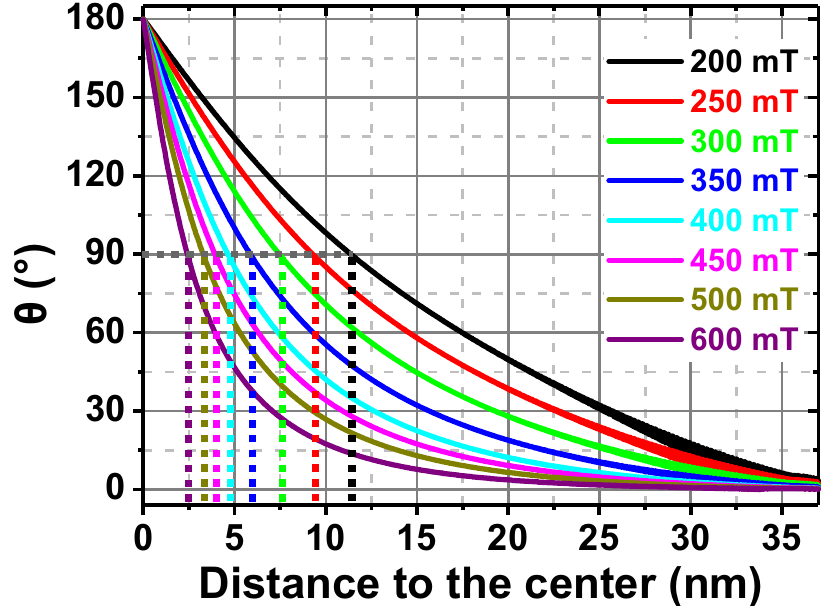}
	\end{center}
	\caption{\label{fig_S6}
		{\bf Determination of the skyrmion radii.}
		 Magnetization orientation from the center to the edge for magnetic field from 200 to \unit{600}{\milli\tesla} of the calculated skyrmion profiles. Dashed vertical lines correspond to the distance to the center where magnetization is in-plane. This distance is half the skyrmion radius  defined throughout this paper.}
\end{figure}

As indicated in the main text, any stray field of the tip modifies the skyrmion structure or laterally moves it during scanning. Therefore, to determine the dependency of the skyrmion shape with magnetic field (Fig. 5b in the main text), a bare tungsten tip was used to image the skyrmion with the TAMR. In oder to take advantage of the high sensitivity of TAMR to distinguish between in-plane and out-of-plane oriented area, the definition of the skyrmion radius used here defer from the one use in \cite{Bocdanov-1994aa}. We measured the diameter of the dark ring observed in the TAMR $dI/dU$ maps (inset Fig. 5b - main text). The theoretical radius was obtained on the same way (green dot - Fig. 5b - main text). \ref{fig_S6} show the theoretical skyrmion profiles of the magnetization orientation from the center to the edge for magnetic field from 200 to \unit{600}{\milli\tesla}. The skyrmion radius was determine by calculating the distance between the two opposite in-plane oriented sections i.e. two times the distance to the center represented by the vertical dashed lines.

{\bf S7 Dipole-dipole interaction in Co/Ru(0001)}

We have estimated the strength of the dipole-dipole interaction. The dipole-dipole energy density is given by:

\begin{eqnarray}
E=\frac{\mu_0\mu_s^2}{4 N \pi} \sum_{i,j=1}^N \frac{ \left( \mathbf{m}_i \cdot\mathbf{m}_j \right) r_{ij}^2 -3 \left( \mathbf{m}_i \cdot \mathbf{r}_{ij} \right) \left( \mathbf{m}_j \cdot \mathbf{r}_{ij} \right)}{ r_{ij}^5 },
\end{eqnarray}

\noindent where $N=\sqrt{3}S/2a^2$ is the number of magnetic moments, $S$ is the surface and $a$ is the lattice parameter ($a=0.27$ nm), $\mu_0$ is the magnetic permeability of vacuum ($\mu_0=5.788.10^{-5}$ eV.T${}^{-1}$), $\mu_s$ is the magnetic moments of Co ($\mu_s=1.8$~$\mu_B$), $\mathbf{m}_i$ the magnetization unit vector at position $\mathbf{r}_i$ and $\mathbf{r}_{ij}= \mathbf{r}_{j} - \mathbf{r}_{i}$.
To quantify the energy density originating from the dipole-dipole interaction, we compare the situation where the magnetization points in different direction for a hexagonal domain of size $270 \times 11.7$ nm${}^2$ with closed and open boundary conditions.

We start by calculating the total energy of the spin-spiral which was obtained from our DFT calculations. When the dipole-dipole energy is included, this spin spiral configuration is the ground state. If we consider homogeneously magnetized states, as expected, the out-of-plane magnetized domain is the maximum of energy. The dipole-dipole energy difference between the out-of-planed and the in-planed magnetized domain is $70 \, \mu$eV/Co and $72 \, \mu$eV/Co with open and closed boundary conditions, respectively and favors an in-plane magnetization .

\newpage
\bibliography{citations2}
\bibliographystyle{nature}
\newpage